\documentclass[12pt]{article}
\usepackage{fullpage} \usepackage{authblk}

\usepackage[round]{natbib} \usepackage{amsmath,amssymb,amsthm} \usepackage{bbold}
\usepackage{graphicx} \usepackage{fancybox} \usepackage{booktabs} 
\usepackage{color}  

\usepackage[nolists,nomarkers,fighead,tabhead]{endfloat}

\usepackage[lofdepth,lotdepth]{subfig}

\newcommand{\E}{\mathbb{E}}

\newcommand{\argmax}{\mathop{\mathrm{argmax}}\limits} 
 \newcommand{\diag}{\mathrm{diag}}
\newcommand{\la}{\langle} \newcommand{\ra}{\rangle}
\newcommand{\ol}{\overline} \newcommand{\wh}{\widehat} \newcommand{\wt}{\widetilde}

\usepackage{bm}

\theoremstyle{break}
 \newtheorem{thm}{Theorem}
 
\newtheorem{prop}[thm]{Proposition} 

\allowdisplaybreaks[1] 

\begin{document}

\title{Mixture regression for observational data, with application to functional regression models}
\author{Toshiya Hoshikawa}
\affil{IMJ Corporation}

\maketitle

\begin{abstract}
  In a regression analysis, suppose we suspect that there are several heterogeneous groups 
  in the population that a sample represents.
  Mixture regression models have been applied to address such problems.
  By modeling the conditional distribution of the response given the covariate as a mixture, 
  the sample can be clustered into groups and the individual regression models for the groups can be
  estimated simultaneously.
  This approach treats the covariate as deterministic so that the covariate carries no information as to 
  which group the subject is likely to belong to.
  Although this assumption may be reasonable in experiments where the covariate is completely determined by 
  the experimenter,
  in observational data the covariate may behave differently across the groups.
  Thus the model should also incorporate the heterogeneity of the covariate, which allows us to estimate 
  the membership of the subject from the covariate.
  
  In this paper, we consider a mixture regression model where the {\it joint} distribution of the response
  and the covariate is modeled as a mixture.
  Given a new observation of the covariate, this approach allows us to compute the posterior probabilities
  that the subject belongs to each group.
  Using these posterior probabilities, the prediction of the response can adaptively use the covariate.
  We introduce an inference procedure for this approach
  and show its properties concerning estimation and prediction.
  The model is explored for the functional covariate as well as the multivariate covariate.
  We present a real-data example where our approach outperforms the traditional approach,
  using the well-known Berkeley growth study data.
\end{abstract}
{\bf Key words:} clustering, functional data, membership, mixture regression, 
observational data, posterior, prediction

\section{Introduction}
In a regression analysis, suppose we suspect that there are several heterogeneous groups 
in the population that a sample represents.
Mixture regression models have been applied to address such problems [\citet{DC88}, \citet{MP00}].
It is assumed in mixture regression that, given a $p$-dimensional covariate $X$ whose subject belongs
to the $k$th group,
the conditional mean of the response $Y$ is related to the linear function of $X$
through a link function $h$ in the format of $h\{\E(Y|X, \delta_{k}=1)\} = \alpha_{k} + \beta_{k}^{T}X$,
where $\delta_{k}$ is the membership variable that returns one if the subject belongs to the $k$th group and 
zero otherwise.
For simplicity, we focus on the normal, identity link model where the conditional density is given by
\begin{align*}
  f_{Y|X, \delta_{k}=1}(y|x) = \varphi(y;\alpha_{k}+\beta_{k}^{T}x,\sigma_{k}^{2}),
\end{align*}
where $\varphi(\cdot;\mu,\sigma^{2})$ is the normal density with mean $\mu$ and variance $\sigma^{2}$.
The EM algorithm can be used to compute the maximum likelihood estimator (MLE),
as often done for finite mixture models [\citet{MP00}].
Information criteria such as Aakaike Information Criterion (AIC) or Bayesian Information Criterion (BIC)
can be used to estimate the number of groups.
\citet{NST07} introduced a modified AIC that is tailored for mixture regression models.

Recently, \citet{YFL11} introduced a mixture regression model where the covariate is given by functional data.
They conducted a real-data analysis and claimed that the mixture regression
approach works better than the (usual) linear regression approach in terms of prediction.
We reconsider this analysis and show that the mixture regression approach works no better than
the linear regression approach when the membership of a new observation is not available.
For the overview on functional data analysis,
readers may refer to excellent monographs written by \citet{RS02,RS05} and \citet{FV06}.

The mixture regression model introduced above treats the covariate as deterministic 
or its distribution as invariant across the groups.
Thus the covariate carries no information as to which group the subject is likely to belong to.
Consider the prediction of the response from a new observation of the covariate;
the best we can do is to take the average of the linear predictors over the groups with certain fixed weights.
Although this assumption may be reasonable in experiments where the covariate is determined 
in a completely deterministic way,
in observational data the covariate may behave differently across the groups.
Thus the model should incorporate the heterogeneity of the covariate as well
so that we can estimate the membership of the subject from the covariate.

In this paper, we introduce a mixture regression model, where the {\it joint} distribution of the response 
and the covariate is modeled as a mixture.
In particular, we assume that the joint density of $X$ and $Y$ is given by
\begin{align*}
  f_{Y,X|\delta_{k}=1}(y,x) = \varphi(y;\alpha_{k}+\beta_{k}^{T}x,\sigma_{k}^{2})\varphi(x,\mu_{k},\Sigma_{k}).
\end{align*}
This is a generalization of the traditional mixture regression model;
when the covariate distribution is identical across the groups, 
this model becomes equivalent to the traditional model.
Our new approach allows the covariate to behave differently across the groups as its marginal distribution
becomes a mixture.
This covariate heterogeneity allows us to compute the posterior probabilities that the subject belongs 
to each group;
using these posterior probabilities, the prediction of the response can adaptively use the covariate.
This assumption is particularly reasonable in functional data analysis; in many practical situations,
functional data appears in observational studies.
We introduce one of such examples in Section \ref{sec:data analysis}.

The rest of the paper proceeds as follows.
In Section \ref{sec:model}, we explore our new approach in more details and introduce an inference procedure.
We first consider the multivariate covariate model, followed by the functional covariate model.
Furthermore, we introduce a couple of simple but very effective ways to extend the model 
to improve the prediction performance; these tricks are used in Section \ref{sec:data analysis}.
Section \ref{sec:properties} discusses the properties of the estimator and the predictor in the joint mixture
regression model.
In Section \ref{sec:simulation}, we explore the properties of our new approach by simulation studies.
Section \ref{sec:data analysis} presents a real-data analysis where we show how our approach can improve 
the prediction performance from the traditional approach, by using the well-analyzed Berkeley growth 
study data.
Finally, we conclude the paper with some remarks in Section \ref{sec:discussion}.

\section{Joint mixture regression}
\label{sec:model}
\subsection{The multivariate covariate model}
Let us first consider the model for the multivariate covariate.
Denote the response by $Y$ and the $p$-dimensional covariate by $X$.
We consider the model where the joint distribution of $(Y,X)$ is a mixture 
whose density is given by
\begin{align}
  \label{eq:model}
  f(y, x)
  = \sum_{k=1}^{K}\pi_{k}\varphi(y;\alpha_{k}+\beta_{k}^{T}x,\sigma_{k}^{2})
  \varphi(x; \mu_{k},\Sigma_{k}),
\end{align}
where $\varphi(\cdot;\mu,\Sigma)$ is the (multivariate) normal density 
with mean $\mu$ and variance(-covariance matrix) $\Sigma$.
$\alpha_{k}$ and $\beta_{k}$ are respectively the regression intercept and slope 
for the $k$th mixture component.
Within each mixture component, which represents a group, the marginal distribution of the covariate is given
by a normal distribution whose parameters vary across the components.
As noted in Introduction, this model differs from the traditional mixture regression models 
in that the traditional approach does not incorporate the covariate distribution into the model
[\citet{MP00}, \citet{NST07}, \citet{YFL11}].
In particular, the traditional approach is based on the {\it conditional} distribution
while in our approach the {\it joint} distribution is assumed to be a mixture.
We call the former {\it ordinary mixture regression} (OMR) and the latter {\it joint mixture regression}
(JMR).
$(\pi_{1},\dots, \pi_{K})$ are mixing proportions, i.e., $\pi_{k}> 0$ and $\sum_{k}\pi_{k}=1$,
and $K$ is the number of the mixture components.
To avoid identifiability issues, we assume that $K$ is the smallest in the sense that 
there is no expression that has fewer components than $K$ but still retains the identical distribution.
We also treat the parameter space as the quotient space with respect to permutation in axes.

An alternative expression which is equivalent to (\ref{eq:model}) and will
become useful when exploring the functional covariate model in the next section is given by
\begin{align}
  \label{eq:model2}
  \begin{split}
    Y &= \sum_{k=1}^{K}\delta_{k}(\alpha_{k} + \beta_{k}^{T}X + \varepsilon_{k}) \\
    X &= \sum_{k=1}^{K}\delta_{k}X_{k},
  \end{split}
\end{align}
where $\Delta=(\delta_{1},\dots,\delta_{K})$ follows a multinomial distribution with parameters $n=1$ 
and $(\pi_{1},\dots,\pi_{K})$,
$X_{k}\sim N(\mu_{k}, \Sigma_{k})$, $\varepsilon_{k} \sim N(0, \sigma_{k}^{2})$,
and $X_{k}$, $\delta_{k}$ and $\varepsilon_{k}$ are jointly independent.
$\Delta$ is often called the membership vector, which indicates the group to which the subject belongs.
Note that we observe $X$ but not $\Delta$, i.e., the membership.

Let $(x_{1},y_{1}), \dots,(x_{n},y_{n})$ be $n$ independent observations from (\ref{eq:model}), or equivalently
(\ref{eq:model2}).
For a fixed positive integer $K$, we estimate the parameters by maximizing the log-likelihood
\begin{align*}
  \ell_{n}(\Psi;y,x) = \sum_{i=1}^{n}\log\Big\{\sum_{k=1}^{K}\pi_{k}\varphi(y_{i};\alpha_{k}+\beta_{k}^{T}x_{i},
  \sigma_{k}^{2})\varphi(x_{i}; \mu_{k},\Sigma_{k})\Big\},
\end{align*}
where the parameters to be estimated are
\begin{align*}
  \Psi := \{\pi_{1},\dots,\pi_{K-1},\alpha_{1},\beta_{1},\sigma_{1}^{2},\mu_{1},\Sigma_{1},\dots,
  \alpha_{K},\beta_{K},\sigma_{K}^{2},\mu_{K},\Sigma_{K}\}.
\end{align*}
As commonly used in finite mixture models, the EM algorithm can be used to compute the MLE,
where $\Delta$ is treated as missing values.
The explicit formula of the algorithm is given in Appendix \ref{sec:em}.

The number of the mixture components $K$ can be estimated through Bayesian Information Criterion (BIC), i.e.,
\begin{align*}
  \wh K = \argmax_{K}\Big\{\max_{\Psi}\ell_{n}(\Psi;y,x) - \frac{|\Psi|}{2}\log n\Big\},
\end{align*}
where $|\Psi|$ is the number of the parameters.
Under some regularity conditions such as the compactness of the parameter space,
BIC provides a consistent estimator for $K$.
For the details, see \citet{Keribin00}.

To predict the response, it is natural to use the empirical best predictor.
Given a new observation of the covariate $X$, the posterior probabilities of the membership are given 
by $\E(\delta_{k}|X)$.
The best predictor of $Y$ is then given by
\begin{align*}
  \E(Y|X) &= \sum_{k=1}^{K}p_{k}(X)(\alpha_{k} + \beta_{k}^{T}X),
\end{align*}
where
\begin{align*}
  p_{k}(X) := \E(\delta_{k}|X) =  \frac{\pi_{k}\varphi(X; \mu_{k},\Sigma_{k})}
  {\sum_{k=1}^{K}\pi_{k}\varphi(X; \mu_{k},\Sigma_{k})}.
\end{align*}
Note that the averaging weights for the conditional group means are given by the function of $X$, so that
the prediction of the response can adaptively use the covariate information to adjust the weights.
In OMR, in contrast, the best predictor is given by the weighted average with $\pi_{k}$ used as the fixed
weights, i.e., $\sum_{k=1}^{K}\pi_{k}(\alpha_{k}+\beta_{k}^{T}X)$.
Intuitively speaking, the more separated the covariate distribution is across the groups, 
the better the prediction performance of the JMR approach will be by adaptively changing the weights,
compared to the OMR approach.

The sample can be clustered by assigning a subject to the group whose empirical posterior is the largest.
For instance, the $i$th subject is assigned to the group
\begin{align*}
  \argmax_{k}\frac{\wh\pi_{k}\varphi(y_{i};\wh\alpha_{k} + \wh\beta_{k}^{T}x_{i},
    \wh\sigma_{k}^{2})\varphi(x_{i}; \wh\mu_{k},\wh\Sigma_{k})}
  {\sum_{k=1}^{K}\wh\pi_{k}\wh\varphi(y_{i};\wh\alpha_{k} + \wh\beta_{k}^{T}x_{i},
    \wh\sigma_{k}^{2})\varphi(x_{i}; \wh\mu_{k},\wh\Sigma_{k})}.
\end{align*}

In Section \ref{sec:simulation}, JMR and OMR are numerically compared.
We show that not only JMR performs better when the covariate distribution is heterogeneous across the groups,
OMR possesses little advantage over JMR even when the covariate distribution is homogeneous and OMR is 
the correctly specified approach.
Furthermore, it is shown that OMR works no better than fitting a linear regression model 
in terms of the prediction performance.
We further confirm these properties with a real data in Section \ref{sec:data analysis}.

\mbox{} \\

\noindent
{\bf Remark:} One may claim that the assumption of the normality for the covariate distribution 
is too restrictive.
Another way to see JMR is to treat it as flexible approximation to the unknown population distribution
by a mixture that can account for the heterogeneity of the covariate distribution as well.
Under this interpretation, the number of the mixture components work as a tuning parameter 
[\citet{GW00}, \citet{GV01}].
In this paper, we leave this aspect of the problem aside and assume that the population model is
(\ref{eq:model}) and the number of the mixture components have a physical meaning.

\subsection{The functional covariate model}
Let us now extend the joint mixture regression (JMR) model to incorporate the functional covariate 
into the model.
Replacing the multivariate covariate in (\ref{eq:model2}) with a random function $X(t)\in L^{2}[0,1]$
(for simplicity, assume its domain is $[0,1]$), we have
\begin{align}
  \label{eq:model3}
  \begin{split}
    Y &= \sum_{k=1}^{K}\delta_{k}(\alpha_{k} + \la\beta_{k},X\ra + \varepsilon_{k})\\
    X &= \sum_{k=1}^{K}\delta_{k}X_{k},
  \end{split}
\end{align}
where $\beta_{k}(t)\in L^{2}[0,1]$.
The inner product is defined by the usual $L^{2}$ inner product, i.e., $\la f,g\ra = \int_{0}^{1} f(t)g(t)dt$.
Let $X_{k}(t)$ be a Gaussian process with mean function $\mu_{k}(t)$ and covariance function $\Gamma_{k}(s,t)$. 
Assume that the covariance function of $X$, say $\Gamma$, allows the eigen-decomposition
\begin{align*}
  \Gamma(s,t) = \sum_{j=1}^{\infty}\lambda_{j}\psi_{j}(s)\psi_{j}(t),
\end{align*}
where $\lambda_{1}\geq\lambda_{2}\geq\dots\geq0$ and $\{\psi_{1},\psi_{2},\dots\}$ forms a complete 
orthonormal basis in $L^{2}[0,1]$ [Mercer's theorem, see \citet{AG75}].
Then, $X$ allows the Karhunen-Lo\`eve decomposition
\begin{align}
  \label{eq:KL-decomp}
  X(t) = \mu(t) + \sum_{j=1}^{\infty}\xi_{j}\psi_{j}(t),
\end{align}
where $\mu(t) = \E X(t) = \sum_{k=1}^{K}\pi_{k}\mu_{k}(t)$ and $\xi_{j} = \la X-\mu, \psi_{j}\ra$
[see \citet{AG75}].
$\xi_{j}$ has mean 0 and variance $\lambda_{j}$, and $\xi_{j}$ and $\xi_{j'}$ are independent for $j\neq j'$;
without the Gaussianity assumption, they are uncorrelated but not necessarily independent.
Plugging (\ref{eq:KL-decomp}) into (\ref{eq:model3}) yields
\begin{align}
  \label{eq:model4}
  Y = \sum_{k=1}^{K}\delta_{k}(a_{k} + \sum_{j=1}^{\infty}b_{kj}\xi_{j} + \varepsilon_{k}),
\end{align}
where $a_{k} = \alpha_{k} + \sum_{j=1}^{\infty}b_{kj}\la\mu,\psi_{j}\ra$ and $b_{kj}=\la\beta_{k},\psi_{j}\ra$.
Note that $\xi_{j}=\sum_{k=1}^{K}\delta_{k}\la X_{k}-\mu,\psi_{j}\ra$ and
$(\la X_{k}-\mu,\psi_{1}\ra, \la X_{k}-\mu,\psi_{2}\ra, \dots)$ is a discrete Gaussian process,
so that $(\xi_{1},\xi_{2},\dots) = \sum_{k=1}^{K}\delta_{k}(\la X_{k}-\mu,\psi_{1}\ra, \la X_{k}-\mu,\psi_{2}\ra,
\dots)$ is a finite mixture of discrete Gaussian processes.
Thus the model (\ref{eq:model4}) can be viewed as generalization of the multivariate model (\ref{eq:model2})
to the infinite-dimensional covariate model.

Unlike the model (\ref{eq:model2}), the problem is now infinite dimensional and we do not directly 
observe $\xi_{j}$.
To reduce the dimensionality we follow the commonly used approach in the functional data analysis literature
[\citet{MS05}, \citet{CH06}, \citet{HH07}, \citet{YFL11}].
With sufficiently large positive integer $M$, 
assume that $\beta_{k}$ can be spanned by $M$ leading eigenfunctions, i.e., 
$\beta_{k}(t) = \sum_{j=1}^{M}b_{kj}\psi_{k}(t)$ for $k=1,\dots,K$.
This assumption turns (\ref{eq:model4}) into
\begin{align}
  \label{eq:model5}
  Y = \sum_{k=1}^{K}\delta_{k}(a_{k} + {b_{k}^{*}}^{T}\xi^{*} + \varepsilon_{k}),
\end{align}
where $b_{k}^{*} = (b_{k1},\dots, b_{kM})^{T}$ and $\xi^{*}=(\xi_{1},\dots,\xi_{M})^{T}$.
This is essentially equivalent to the multivariate covariate model, 
except that $\xi^{*}$ is not directly observable.
We estimate $\xi^{*}$ and use its estimate as a surrogate.

In practice, the functional covariate is not continuously observable; only a finite number of observations 
at discrete points per curve are available.
Suppose there are $n$ realizations $(Y_{1},X_{1}),\dots,(Y_{n},X_{n})$.
The form of the sample available to us is
\begin{align*}
  \{Y_{1},X_{1}(t_{1,1}),\dots,X_{1}(t_{1,m_{1}})\},\dots, \{Y_{n},X_{n}(t_{n,1}),\dots,X_{n}(t_{n,m_{n}})\},
\end{align*}
where $m_{1},\dots,m_{n}$ are the numbers of observation points per curve,
and the sets of the observation points are not necessarily synchronized nor equally discretized.
From these observations, we have to estimate $\xi_{1}^{*},\dots, \xi_{n}^{*}$.
The analysis of the components of the Karhunen-Lo\`eve decomposition, i.e., 
$\lambda_{j}, \psi_{j}, \xi_{ij} \ (i=1, \dots, n, \ j=1, 2, \dots)$, is called functional principal component
analysis (FPCA),
and has been developed for the past two decades by many authors.
To save space, we avoid going into the details on the FPCA techniques, but interested readers
may refer to \citet{YMW05}, \citet{HMW06}, \citet{BHK09}, and the references therein.
The point is that we can estimate $\xi_{1}^{*},\dots, \xi_{n}^{*}$ by using a technique in FPCA.
In Section \ref{sec:data analysis} where we apply the functional covariate model to a real-data analysis,
we follow Ramsay and Silverman's paradigm of mapping a curve onto the space spanned by a finite number of 
basis functions [\citet[2005]{RS02}; an excellent package ``fda'' is available for R and MATLAB].

Given estimates $\wh\xi_{i}^{*}$, we can estimate the parameters in the model (\ref{eq:model5}) 
by maximizing the estimated log-likelihood, where $\xi_{i}^{*}$ is replaced with $\wh\xi_{i}^{*}$
in the true log-likelihood, i.e.,
\begin{align}
  \label{eq:loglik}
  \ell_{n}(\Psi;y,\wh\xi^{*}) = \sum_{i=1}^{n}\log\Big\{\sum_{k=1}^{K}\pi_{k}\varphi(y_{i};a_{k} + 
  {b_{k}^{*}}^{T}\wh\xi_{i}^{*}, \sigma_{k}^{2})\varphi(\wh\xi_{i}^{*}; \mu_{k},\Sigma_{k})\Big\},
\end{align}
where the parameters to be estimated are
\begin{align*}
  \Psi = \{\pi_{1},\dots,\pi_{K-1},a_{1},b_{1}^{*},\sigma_{1}^{2},\mu_{1},\Sigma_{1},\dots,
  a_{K},b_{K}^{*},\sigma_{K}^{2},\mu_{K},\Sigma_{K}\}.
\end{align*}
Finally, the regression slope functions $\beta_{k}$ can be estimated by
\begin{align*}
  \wh\beta_{k}(t) = \sum_{j=1}^{M}\wh b_{kj}\wh\psi_{k}(t),
\end{align*}
which is a consistent estimator (Section \ref{sec:properties}).

The procedures for prediction and clustering are similar to the multivariate covariate model.
In Section \ref{sec:data analysis}, we apply the functional covariate model to the Berkley growth study data.

\subsection{Tricks to improve the model}
\label{sec:incorp}
In this section, we introduce two ways to improve the model.
First, recall that the heterogeneity of the covariate distribution plays a crucial role in prediction
because it allows us to estimate the membership from the covariate.
The more separated the covariance distribution is across the groups, the better the prediction performance 
will be, because it becomes easier to differentiate the membership.
It is well known in the functional data analysis literature that sometimes higher order derivatives,
$X', X'', \cdots$, shows a clearer difference by group than the original $X$.
A famous example given in \citet{RS05} is a velocity or acceleration curve, which shows much clearer
distinction between gender than a growth curve.
One way to incorporate, say $X'$, into the model is to apply integration by parts to (\ref{eq:model3}),
which yields
\begin{align*}
  Y &= \sum_{k=1}^{K}\delta_{k}(\alpha_{k} + \la\beta_{k},X\ra + \varepsilon_{k})\\
 &= \sum_{k=1}^{K}\delta_{k}(\alpha_{k} - \gamma_{k}(1)X(1) + \la\gamma_{k},X'\ra + \varepsilon_{k}),
\end{align*}
where $\gamma_{k}(t) = -\int_{0}^{t}\beta_{k}$.
These two expressions are identical in theory, but in practice the latter may perform better 
if $X'$ is more distinguishable by group than $X$.
As the constraint between the regression coefficient of $X(1)$ and the regression slope function of $X'$
is inconvenient to estimate the model,
we may avoid it by treating the regression coefficient as a free parameter, i.e.,
\begin{align*}
  Y &= \sum_{k=1}^{K}\delta_{k}(\alpha_{k} + \zeta_{k} X(1) + \la\gamma_{k},X'\ra + \varepsilon_{k}),
\end{align*}
where $\zeta_{k}$ is a free parameter.
This model includes (\ref{eq:model3}) as a submodel. 

Another way to extend the model (\ref{eq:model3}) is to allow two kinds of covariates:
one that behaves similarly across the groups or is deterministic,
and the other that behaves differently across the groups---sometimes we know beforehand that a certain
covariate, say $Z$, has an invariant distribution or is deterministic such as covariates in experiments.
Adding $Z$ to the model (\ref{eq:model2}), it becomes
\begin{align*}
  Y &= \sum_{k=1}^{K}\delta_{k}(\alpha_{k} + \zeta_{k}^{T}Z + \beta_{k}^{T}X + \varepsilon_{k}).
\end{align*}
Under this model, the inference should be based on the conditional distribution given $Z$ 
so that we can exclude unnecessary parameters from the model that does not contribute to the membership 
estimation.
The EM algorithm can be straightforwardly modified to accommodate this model.
These two extensions are simple, yet very effective to improve the prediction performance, as demonstrated
in Section \ref{sec:data analysis}.

\section{Theoretical properties}
\label{sec:properties}
\subsection{Consistency of the MLE in the functional covariate model}
The maximizer of (\ref{eq:loglik}) does not coincide with the actual maximum likelihood estimator 
because $\xi^{*}$ is replaced with $\wh\xi^{*}$.
Fortunately, the theory in \citet{YFL11} straightforwardly applies to the current problem as well;
under some regularity conditions the maximum likelihood estimator in (\ref{eq:loglik})
is still consistent.
We assume that $\wh\xi_{k}, \wh\psi_{k}, k=1,\dots,M,$ are obtained by using the technique
in \citet{YMW05}.

\begin{prop}
  Assume that the population model is (\ref{eq:model5}) and the assumptions A1 to A4 in \citet{YFL11} hold.
  For any fixed compact set containing the true parameter $\Psi$ as an interior point, 
  let $\wh\Psi$ be the maximizer of (\ref{eq:loglik}) over the compact set.
  Then, $\wh\Psi$ converges to $\Psi$ in probability.
  Furthermore, $\wh\beta_{k}, k=1, \dots, K$ is uniformly consistent, i.e., 
  $\sup_{t\in [0,1]}|\wh\beta_{k}(t)-\beta_{k}(t)|$ converges to 0 in probability,
\end{prop}
There are two aspects of the model that involve the proof: the proximity of $\wh\xi_{k}$ to $\xi_{k}$ and
the local behavior of the log-likelihood function
\begin{align*}
  \ell(\Psi; y, \xi) = \log\Big\{\sum_{k=1}^{K}\pi_{k}\varphi(y;a_{k}+b_{k}^{T}\xi,\sigma_{k}^{2})
  \varphi(\xi_{k};\mu_{k},\Sigma_{k})\Big\}.
\end{align*}
Since the covariate distribution continues to satisfy the assumptions in \citet{YFL11},
the conditions concerning the first aspect are satisfied.
On the other hand, since the likelihood function in JMR has a different form than the one in OMR 
(\citet{YFL11} considered the functional covariate model for OMR),
we need to check whether the current likelihood still retains appropriate local behavior.
In Appendix \ref{sec:proof1}, we verify that the log-likelihood function in JMR also satisfies the 
regularity conditions.

\subsection{Asymptotic mean squared prediction error}
As mentioned before, it is essential to estimate the membership from the covariate in order to predict 
the response well.
In this section, we compare the asymptotic mean square prediction error (MSPE) between JMR and OMR.
Recall that we predict the response by the empirical best predictor
\begin{align}
  \label{eq:prediction}
  \wh Y &= \sum_{k=1}^{K}\wh p_{k}(X)(\wh\alpha_{k} + \wh\beta_{k}^{T}X),
\end{align}
where in JMR
\begin{align}
  \label{eq:partpost}
  \wh p_{k}(X) = \frac{\wh\pi_{k}\varphi(X; \wh\mu_{k},\wh\Sigma_{k})}
  {\sum_{k=1}^{K}\wh\pi_{k}\varphi(X; \wh\mu_{k}, \wh\Sigma_{k})},
\end{align}
while in OMR $\wh p_{k}(X) = \wh\pi_{k}$.
As seen in the last section, the MLE is consistent under the JMR model.
Now, suppose that the population model is the JMR model, but the MLE is obtained by applying the OMR approach.
It may be reasonable to suspect that the resulting MLE is no longer consistent.
However, several numerical explorations that the author conducted including those given in 
Section \ref{sec:simulation} suggest that the MLE obtained by applying the OMR approach is also consistent.
(We have not been successful in proving either this conjecture is true or false.)
We will come back to this point again in the next section.
In the following, we consider two cases concerning the OMR approach: one where the parameters are 
consistently estimated, and the other where the parameters are not consistently estimated.

Consider the multivariate covariate model (\ref{eq:model2}).
For simplicity, let $\alpha_{1}=\dots=\alpha_{K}=0$ and use the inner-product notation, i.e.,
$\la x, y\ra = x^{T}y$.
If the covariate distribution varies across the groups, (\ref{eq:prediction}) provides the smallest asymptotic
MSPE among any possible predictors because it is the MSPE of the population conditional mean.
The asymptotic MSPE is then given by the error variance,
$\varSigma := \sum_{k=1}^{K}\pi_{k}\sigma_{k}^{2}$, plus
\begin{align}
  \label{eq:mspe1}
  \E\Big[\sum_{k=1}^{K}\E(\delta_{k}|X)\Big(\sum_{\ell=1}^{K}\E(\delta_{\ell}|X)
    \la\beta_{k}-\beta_{\ell},X\ra\Big)^{2}\Big].
\end{align}
If we use $\wh p_{k}(X)=\wh\pi_{k}$, the asymptotic MSPE becomes $\varSigma$ plus
\begin{align}
  \label{eq:mspe2}
  \E\Big[\sum_{k=1}^{K}\E(\delta_{k}|X)\Big(\sum_{\ell=1}^{K}\E(\delta_{\ell})\la\beta_{k}-\beta_{\ell},
    X\ra\Big)^{2}\Big],
\end{align}
which is strictly greater than (\ref{eq:mspe1}) unless $\E(\delta_{k}|X)=\E(\delta_{k})$ for all $k$,
nor $\la\beta_{1},X\ra = \dots = \la\beta_{K},X\ra$ almost surely.
The former case implies that the covariate distribution is invariant across the groups, which contradicts 
the assumption.
In the latter case, (\ref{eq:mspe1}) = (\ref{eq:mspe2}) = 0; but in this case the ability to differentiate
the group is not necessary because there is no harm by assuming a wrong group.

Now, suppose that the MLE is asymptotically biased and $\wh\beta_{k}$, $\wh\pi_{k}$ converges to 
some $\beta_{k}^{*}$, $\pi_{k}^{*}$, respectively.
Then, the asymptotic MSPE becomes $\varSigma$ plus
\begin{align}
  \label{eq:mspe3}
  \sum_{k=1}^{K}\pi_{k}\E\Big[\Big\la\sum_{\ell=1}^{K}\pi_{\ell}^{*}(\beta_{k}-\beta_{\ell}^{*}),
  X_{k}\Big\ra^{2}\Big].
\end{align}
This quantity is in fact greater than (\ref{eq:mspe2}) at least when 
$\E(X_{1}X_{1}^{T}) = \dots = \E(X_{K}X_{K}^{T})$.
Without this assumption, the effect of the bias is rather involved as it is easy to create an example
where (\ref{eq:mspe3}) is smaller than (\ref{eq:mspe2}).
The proofs are given in Appendix \ref{sec:proof2}.

\section{Simulation study}
\label{sec:simulation}
This section illustrates how the JMR approach works in comparison to alternative methods.
We generate a sample from the two-dimensional covariate, two-group model
\begin{align*}
  Y &= \delta_{1}(\alpha_{1} + \beta_{1}^{T}X + \varepsilon_{1}) 
  + (1-\delta_{1})(\alpha_{2} + \beta_{2}^{T}X + \varepsilon_{2}) \\
  X &= \delta_{1}X_{1} + (1-\delta_{1})X_{2},
\end{align*}
where the mixing proportion is $(\pi_{1}, \pi_{2}) = (0.6, 0.4)$ and the error variances are both $0.3^{2}$.
The training sample size is considered for 100 and 300, and the testing sample size is 500.
The other parameters---regression coefficients, covariate means, and covariate variance-covariance 
matrices---are determined to construct the following four scenarios:
\begin{enumerate}
\item $X$ and $Y$ are both well separated by group.
\item $X$ has the common group means, and $Y$ is well separated by group.
\item $X$ is well separated by group, but $Y$ is not.
\item $X$ has the common cluster distributions, and $Y$ is well separated by group.
\end{enumerate}
\begin{figure}[h]
  \centering
  \subfloat[Scenario 1]{
    \includegraphics[width=0.45\textwidth]{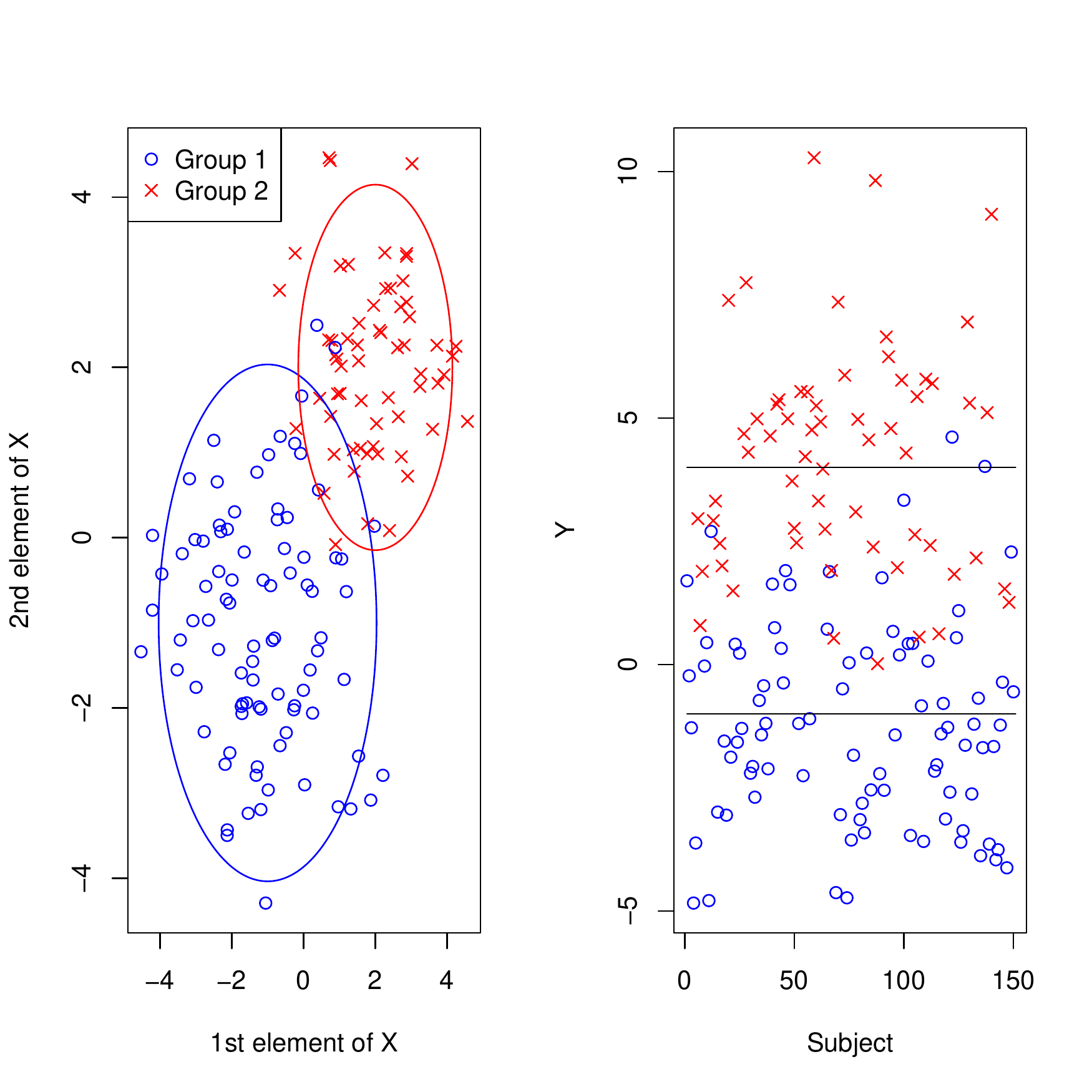}
  }
  \qquad
  \subfloat[Scenario 2]{
    \includegraphics[width=0.45\textwidth]{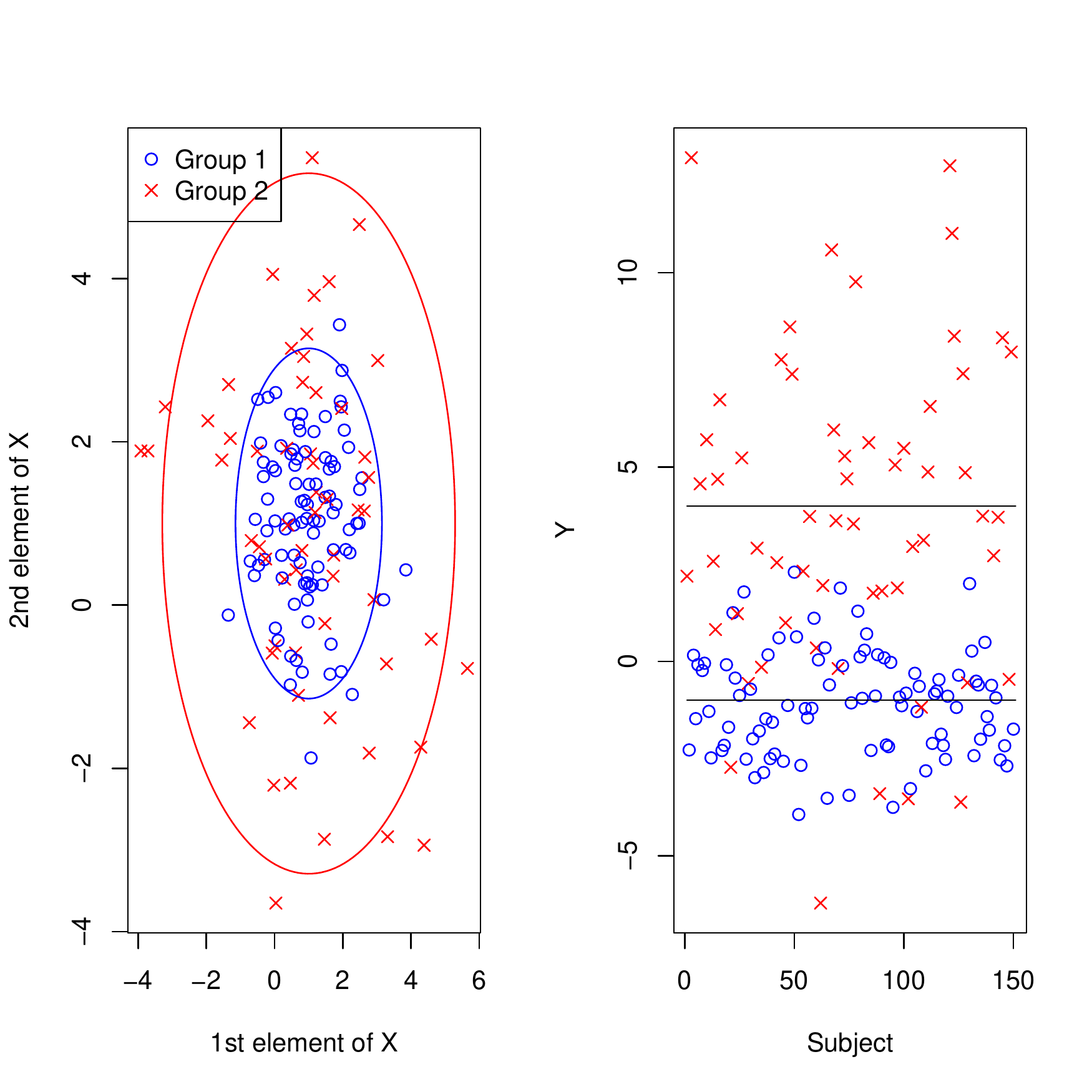}
  }
  \qquad
  \subfloat[Scenario 3]{
    \includegraphics[width=0.45\textwidth]{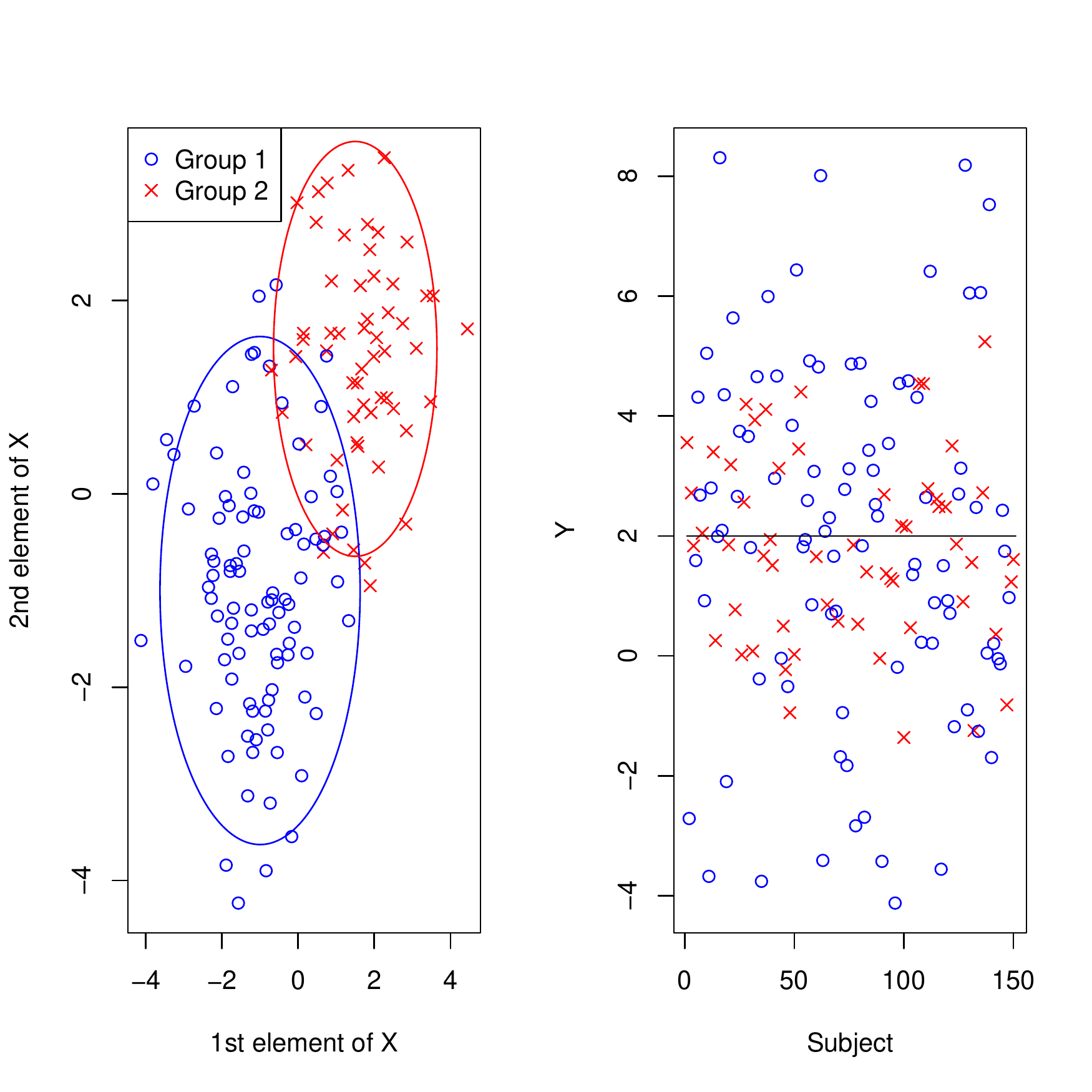}
  }
  \qquad
  \subfloat[Scenario 4]{
    \includegraphics[width=0.45\textwidth]{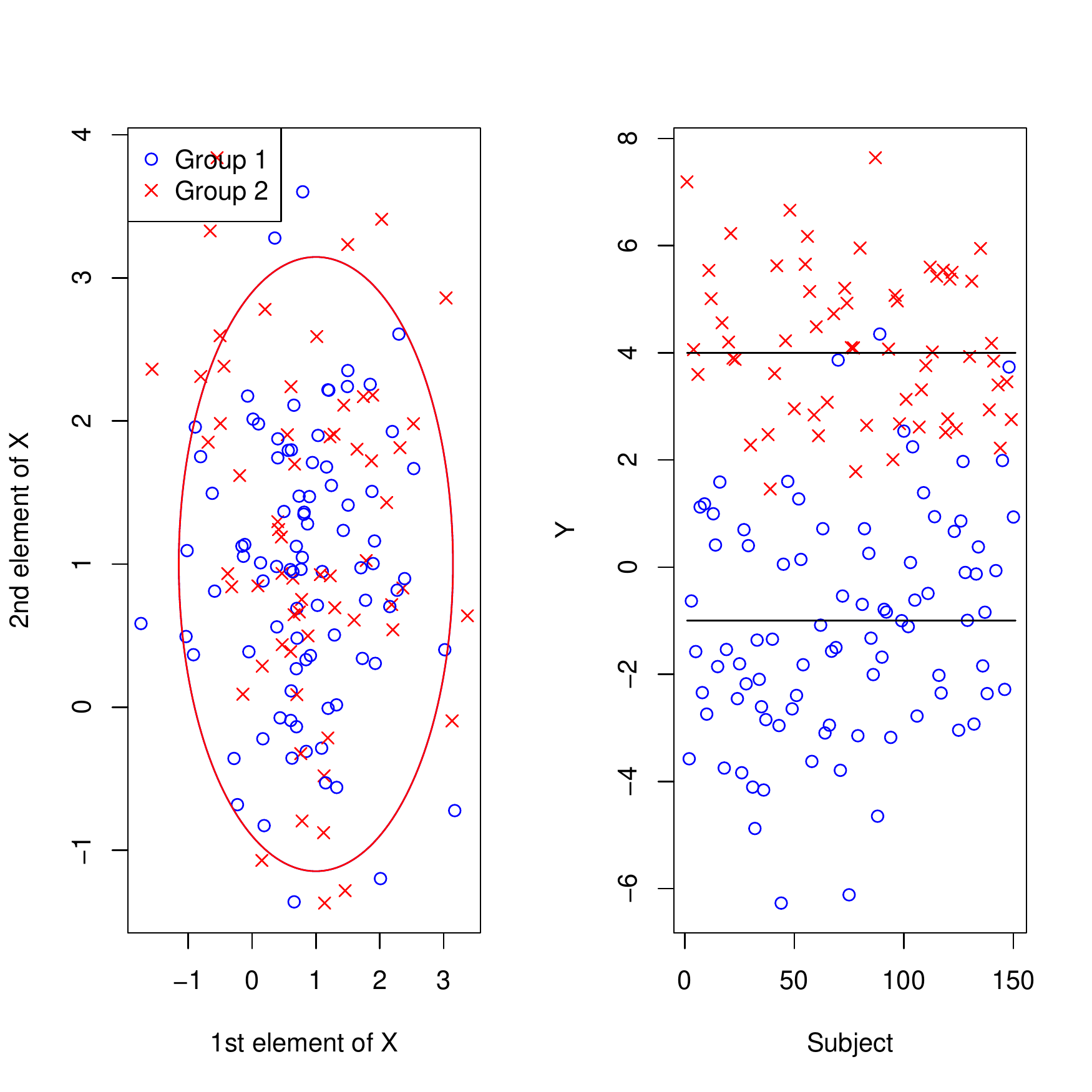}
  }
  \caption{A realization of the training sample of n=150 for each scenario.
    The covariate is plotted in the left figure where the two circles show .95th quantile contours.
    The response is plotted in the right figure where the two horizontal lines
    are the population means.}
  \label{fig:realization}
\end{figure}
Figure \ref{fig:realization} shows a realization of the sample in each scenario.
We calculate the mean squared prediction error (MSPE), the average misclassification rate (MCR),
and the mean squared error (MSE) for part of the parameters over 500 iterations.
We compare JMR to three alternative approaches:
linear regression by ordinary least squares (OLS), ordinary mixture regression (OMR),
and the two-step model-based clustering approach (MBC).
MCR cannot be computed for OLS as it does not cluster a sample.
MBC works as follows.
First, it clusters a sample into two groups by fitting a mixture of normal to the covariate (model-based
clustering); these two groups are used to compute the MCR.
Second, in each cluster the linear regression model is estimated by fitting OLS.
To predict the response, it uses the weighted average of the linear predictors from the two estimated 
linear models with the posterior probabilities calculated from a new observation of the covariate 
used as weights.
We used Fraley and Raftery's R package ``mclust'' for this approach [\citet{FR02}].
Note that JMR is the correctly specified approach in Scenarios 1--3
while OMR is the correctly specified approach in Scenario 4.
It is well known that the clusters obtained in MBC are not identically-distributed samples of the component 
distributions, so that the estimates based on the resulting clusters are inevitably biased.

\begin{table}[ht]
  \centering
  \subfloat[Scenario 1 (n=100)]{
    \begin{tabular}{rrrrr}
      \toprule
      Method & OLS & OMR & JMR & MBC \\ 
      \midrule
      MSPE & 1.66 & 2.25 & 0.35 & 0.44 \\ 
      MCR &  & .074 & .012 & .057 \\
      \bottomrule
    \end{tabular}
  }
  \qquad \vspace{.2in}
  \subfloat[Scenario 1 (n=300)]{
    \begin{tabular}{rrrrr}
      \toprule
      Method & OLS & OMR & JMR & MBC \\ 
      \midrule
      MSPE & 1.62 & 2.23 & 0.31 & 0.33 \\ 
      MCR &  & .067 & .009 & .042 \\
      \bottomrule
    \end{tabular}
  }
  \qquad
  \subfloat[Scenario 2 (n=100)]{
    \begin{tabular}{rrrrr}
      \toprule
      Method & OLS & OMR & JMR & MBC \\ 
      \midrule
      MSPE & 10.46 & 12.35 & 9.50 & 11.05 \\ 
      MCR & & .022 & .023 & .331 \\
      \bottomrule
    \end{tabular}
  }
  \qquad \vspace{.2in}
  \subfloat[Scenario 2 (n=300)]{
    \begin{tabular}{rrrrr}
      \toprule
      Method & OLS & OMR & JMR & MBC \\ 
      \midrule
      MSPE & 10.11 & 12.21 & 9.22 & 10.43 \\ 
      MCR & & .019 & .022 & .274 \\
      \bottomrule
    \end{tabular}
  }
  \qquad
  \subfloat[Scenario 3 (n=100)]{
    \begin{tabular}{rrrrr}
      \toprule
      Method & OLS & OMR & JMR & MBC \\ 
      \midrule
      MSPE & 4.08 & 4.79 & 1.19 & 1.35 \\ 
      MCR &  & .048 & .063 & .080 \\
      \bottomrule
    \end{tabular}
  }
  \qquad \vspace{.2in}
  \subfloat[Scenario 3 (n=300)]{
    \begin{tabular}{rrrrr}
      \toprule
      Method & OLS & OMR & JMR & MBC \\ 
      \midrule
      MSPE & 3.97 & 4.73 & 0.88 & 0.98 \\ 
      MCR &  & .041 & .013 & .061 \\
      \bottomrule
    \end{tabular}
  }
  \qquad
  \subfloat[Scenario 4 (n=100)]{
    \begin{tabular}{rrrrr}
      \toprule
      Method & OLS & OMR & JMR & MBC \\ 
      \midrule
      MSPE & 8.54 & 8.33 & 8.94 & 8.89 \\ 
      MCR &  & .015 & .016 & .441 \\
      \bottomrule
    \end{tabular}
  }
  \qquad \vspace{.2in}
  \subfloat[Scenario 4 (n=300)]{
    \begin{tabular}{rrrrr}
      \toprule
      Method & OLS & OMR & JMR & MBC \\ 
      \midrule
      MSPE & 8.37 & 8.31 & 8.52 & 8.57 \\ 
      MCR &  & .015 & .016 & .449 \\
      \bottomrule
    \end{tabular}
  }
  \caption{The mean squared prediction error (MSPE) and the average misclassification rate (MCR)
    for four scenarios.}
  \label{tab:sim}
\end{table}

\begin{table}[ht]
  \centering
  \subfloat[Scenario 1 (n=100)]{
    \begin{tabular}{rrrr}
      \toprule
      Method &  OMR & JMR & MBC \\ 
      \midrule
      $\pi_{1} [0.6]$ & .056 & .048 & .074 \\
      $\beta_{12} [1]$ & .063 & .052 & .279 \\
      $\beta_{22} [2]$ & .064 & .049 & .206 \\
      \bottomrule
    \end{tabular}
  }
  \qquad \vspace{.2in}
  \subfloat[Scenario 1 (n=300)]{
    \begin{tabular}{rrrrr}
      \toprule
      Method & OMR & JMR & MBC \\ 
      \midrule
      $\pi_{1} [0.6]$ & .037 & .029 & .039 \\ 
      $\beta_{12} [1]$ & .033 & .028 & .166 \\
      $\beta_{22} [2]$ & .039 & .029 & .150 \\
      \bottomrule
    \end{tabular}
  }
  \qquad
  \subfloat[Scenario 2 (n=100)]{
    \begin{tabular}{rrrrr}
      \toprule
      Method & OMR & JMR & MBC \\  
      \midrule
      $\pi_{1} [0.6]$ & .049 & .051 & .245 \\ 
      $\beta_{12} [-1]$ & .026 & .026 & 2.78 \\
      $\beta_{22} [2]$ & .025 & .026 & 4.54 \\
      \bottomrule
    \end{tabular}
  }
  \qquad \vspace{.2in}
  \subfloat[Scenario 2 (n=300)]{
    \begin{tabular}{rrrrr}
      \toprule
      Method &  OMR & JMR & MBC \\ 
      \midrule
      $\pi_{1} [0.6]$ & .029 & .031 & .188 \\ 
      $\beta_{12} [-1]$ & .015 & .014 & .362 \\
      $\beta_{22} [2]$ & .015 & .015 & .992 \\
      \bottomrule
    \end{tabular}
  }
  \qquad
  \subfloat[Scenario 3 (n=100)]{
    \begin{tabular}{rrrrr}
      \toprule
      Method & OMR & JMR & MBC \\ 
      \midrule
      $\pi_{1} [0.6]$ & .057 & .089 & .096 \\ 
      $\beta_{12} [-2]$ & .164 & .489 & .618 \\
      $\beta_{22} [1]$ & .183 & .584 & .368 \\
      \bottomrule
    \end{tabular}
  }
  \qquad \vspace{.2in}
  \subfloat[Scenario 3 (n=300)]{
    \begin{tabular}{rrrrr}
      \toprule
      Method &  OMR & JMR & MBC \\ 
      \midrule
      $\pi_{1} [0.6]$ & .030 & .032 & .049 \\ 
      $\beta_{12} [-2]$ & .032 & .133 & .429 \\
      $\beta_{22} [1]$ & .030 & .151 & .204 \\
      \bottomrule
    \end{tabular}
  }
  \qquad
  \subfloat[Scenario 4 (n=100)]{
    \begin{tabular}{rrrrr}
      \toprule
      Method &  OMR & JMR & MBC \\ 
      \midrule
      $\pi_{1} [0.6]$ & .051 & .051 & .207 \\ 
      $\beta_{12} [-2]$ & .053 & .054 & 2.50 \\
      $\beta_{22} [1]$ & .053 & .053 & 3.82 \\
      \bottomrule
    \end{tabular}
  }
  \qquad \vspace{.2in}
  \subfloat[Scenario 4 (n=300)]{
    \begin{tabular}{rrrrr}
      \toprule
      Method & OMR & JMR & MBC \\ 
      \midrule
      $\pi_{1} [0.6]$ & .030 & .031 & .176 \\ 
      $\beta_{12} [-2]$ & .028 & .028 & 2.57 \\
      $\beta_{22} [1]$ & .028 & .028 & 2.48 \\
      \bottomrule
    \end{tabular}
  }
  \caption{The square root of the mean squared error for some of the parameters in the four scenarios.
    The true parameter is given in the square brackets next to the symbol.}
  \label{tab:sim_est}
\end{table}

The results are given in Tables \ref{tab:sim} and \ref{tab:sim_est}.
We first look at the prediction performance (Table \ref{tab:sim}).
When the covariate distribution is well separated across the groups (Scenarios 1 and 3),
JMR and MBC outperform the other two methods.
When it is difficult to differentiate the group by the covariate (Scenario 2) or the covariate distribution
is homogeneous (Scenario 4), the overall performance deteriorates and the relative advantage of JMR
reduces.
Note that OMR is not even as good as OLS (Scenarios 1--3), and in Scenario 4 where OMR is the correctly
specified approach, it is not noticeably better than the other approaches.
The reason why OMR is no better than OLS is that computing the average over the linear predictors of
the groups with fixed weights is essentially equivalent to fitting the linear model globally;
then OLS tends to have a smaller variation because it needs to estimate much fewer parameters than OMR.

The results with respect to misclassfication seem a little different.
The clustering performance by JMR is fairly well throughout the scenarios, including Scenario 4.
OMR also works well when the covariate distribution is not much separated (Scenarios 2 and 4).
It is even slightly better than JMR in Scenario 2 where OMR is a misspecified approach.
For scenarios 1 and 3, in contrast, JMR works much better than OMR, though the overall performance of OMR
is still comparable to MBC regardless of the fact that OMR does not take into account the heterogeneity of
the covariate distribution.
This implies that to cluster a sample whose clustering structure lies in the regression structure,
clustering based on the regression is at least as equally important as taking into account the covariate
heterogeneity.

One may wonder whether the differences in the prediction performance in fact attribute to the estimability of 
the group.
Table \ref{tab:sim_est} shows the square root of the mean squared error for some of the parameters.
Note that there is not much difference in the estimation performance between OMR and JMR; in some cases,
OMR is even better than JMR.
As the sample size increases, the MSE of OMR reduces at a similar rate to JMR.
This raises the question as to the consistency of OMR; although OMR is a misspecified approach under the JMR
model, the MLE by OMR may be still consistent for the parameters under the JMR model.
We numerically investigated this conjecture, and the results seem to support it.
(We have not been able to prove analytically whether this claim is true or not.)
Because the parameter estimation by the two methods seems similar, we claim that the difference in the
prediction performance mostly attributes to the estimability of the group.
In other words, whether we can predict the response well largely depends on whether we can estimate the group
that the subject of a new observation is likely to belong to from the covariate.
Otherwise, we cannot expect much beyond simply fitting a linear regression model.


\section{Berkeley growth study, revisited}
\label{sec:data analysis}
In this section, we present a real-data example where the joint mixture regression (JMR) approach improves
the prediction performance of the traditional ordinary mixture regression (OMR) approach.
We use the Berkeley growth study data [\citet{TS54}], which contains the recorded height of boys and girls 
from age 1--18 years old; this is a well-analyzed data set and has been repeatedly used as an illustrating 
example in the functional data analysis literature.
Recent examples using this data set include \citet{CL07}, \citet{TM08}, \citet{HMY09}, and \citet{YFL11}.
The data set contains 39 boys and 54 girls whose height was measured quarterly from 1--2 years old,
annually from 2--8 years old, and biannually from 8--18 years old.
We reconsider the analysis given in \citet{YFL11}, where they considered the problem of predicting the height
at the age of 18 from the height transition during the juvenile period.

\begin{figure}[h]
  \centering
  \subfloat[Boys]{
    \label{fig:gcurve_boys}
    \includegraphics[width=0.45\textwidth]{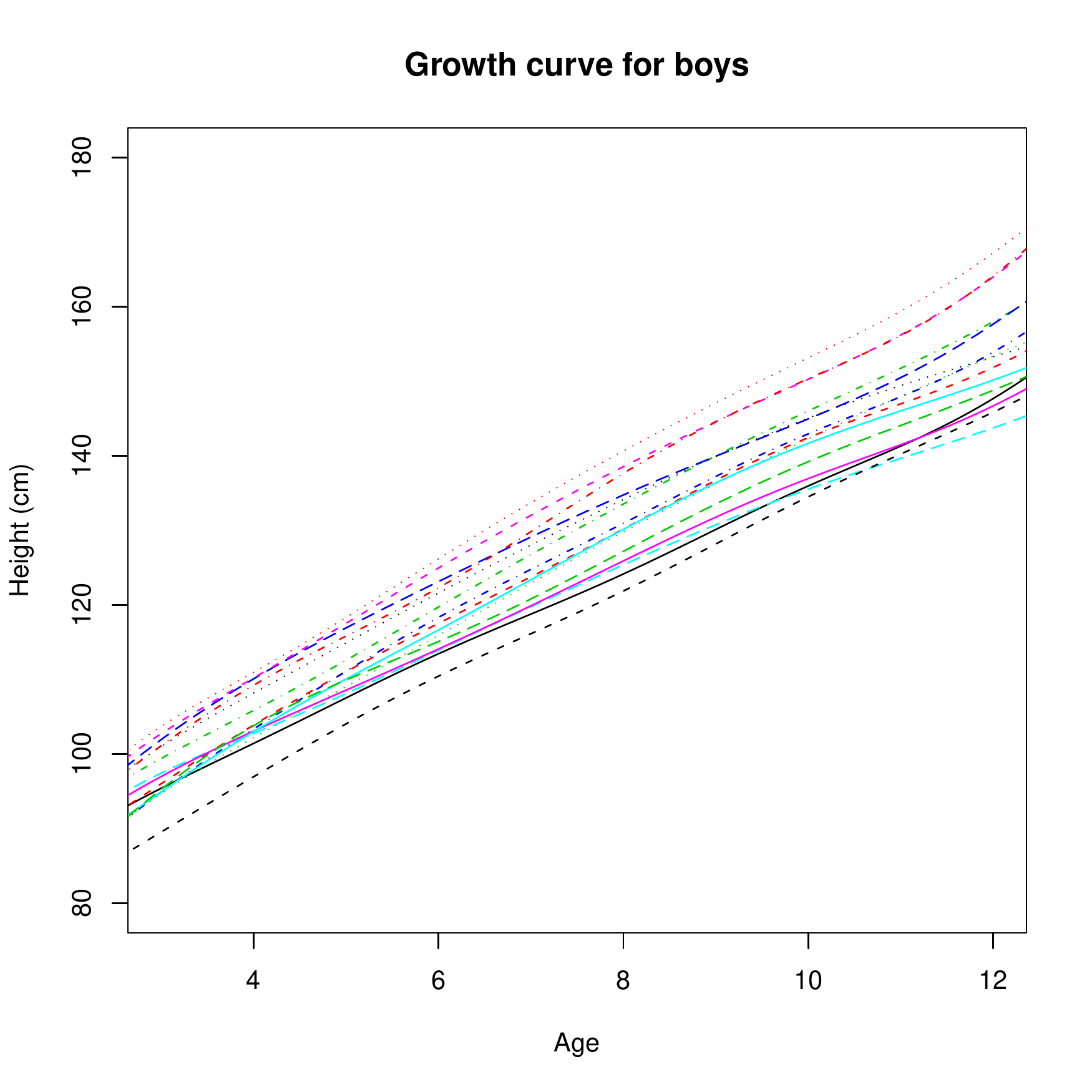}
  }
  \qquad
  \subfloat[Girls]{
    \label{fig:gcurve_girls}
    \includegraphics[width=0.45\textwidth]{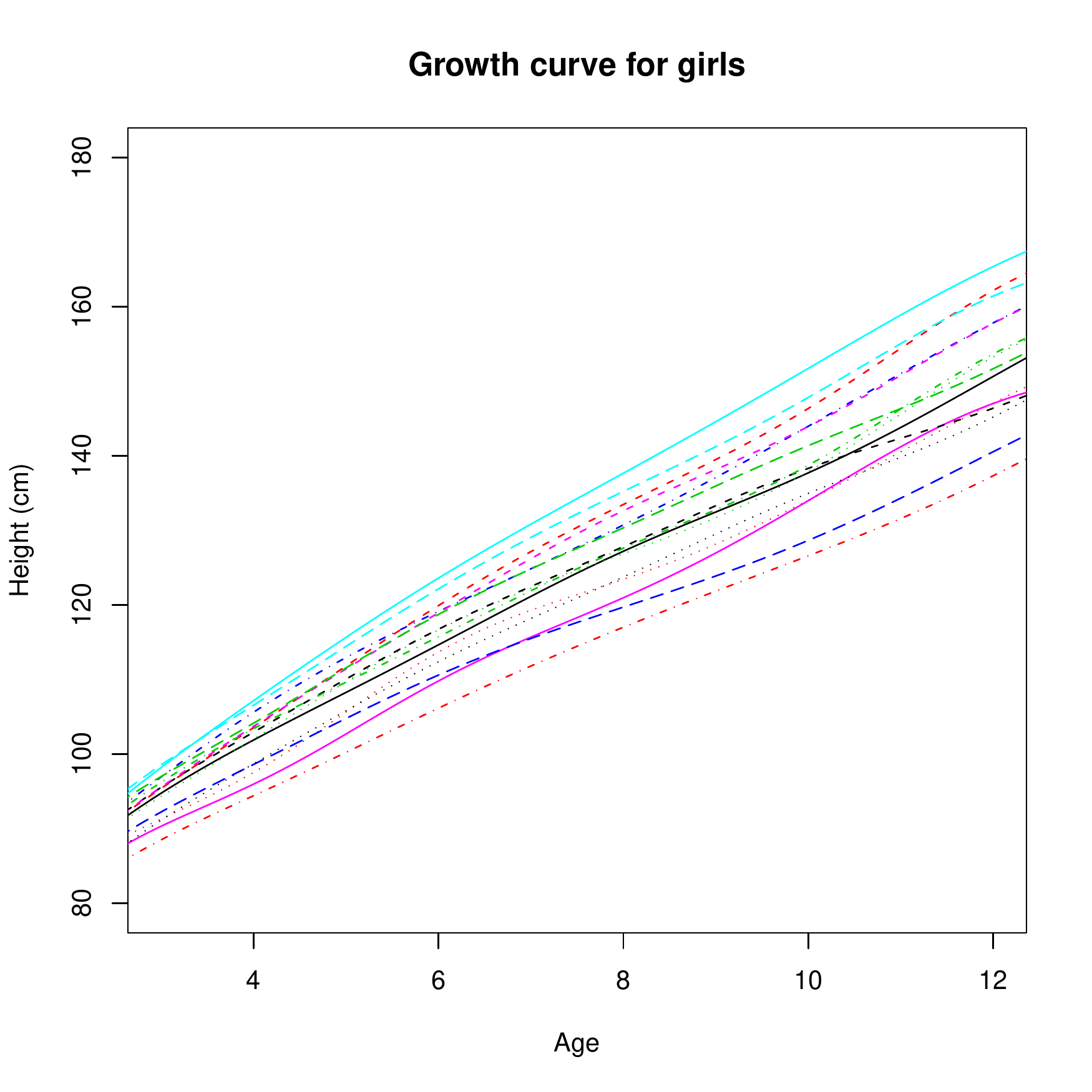}
  }
  \caption{The growth curves for randomly selected 15 boys and 15 girls from age 3-12 years old.
    The curves are obtained by mapping observations onto B-splines of order 5.}
  \label{fig:gcurve}
\end{figure}

\begin{figure}[h]
  \centering
  \subfloat[Model 1]{
    \includegraphics[width=0.53\textwidth]{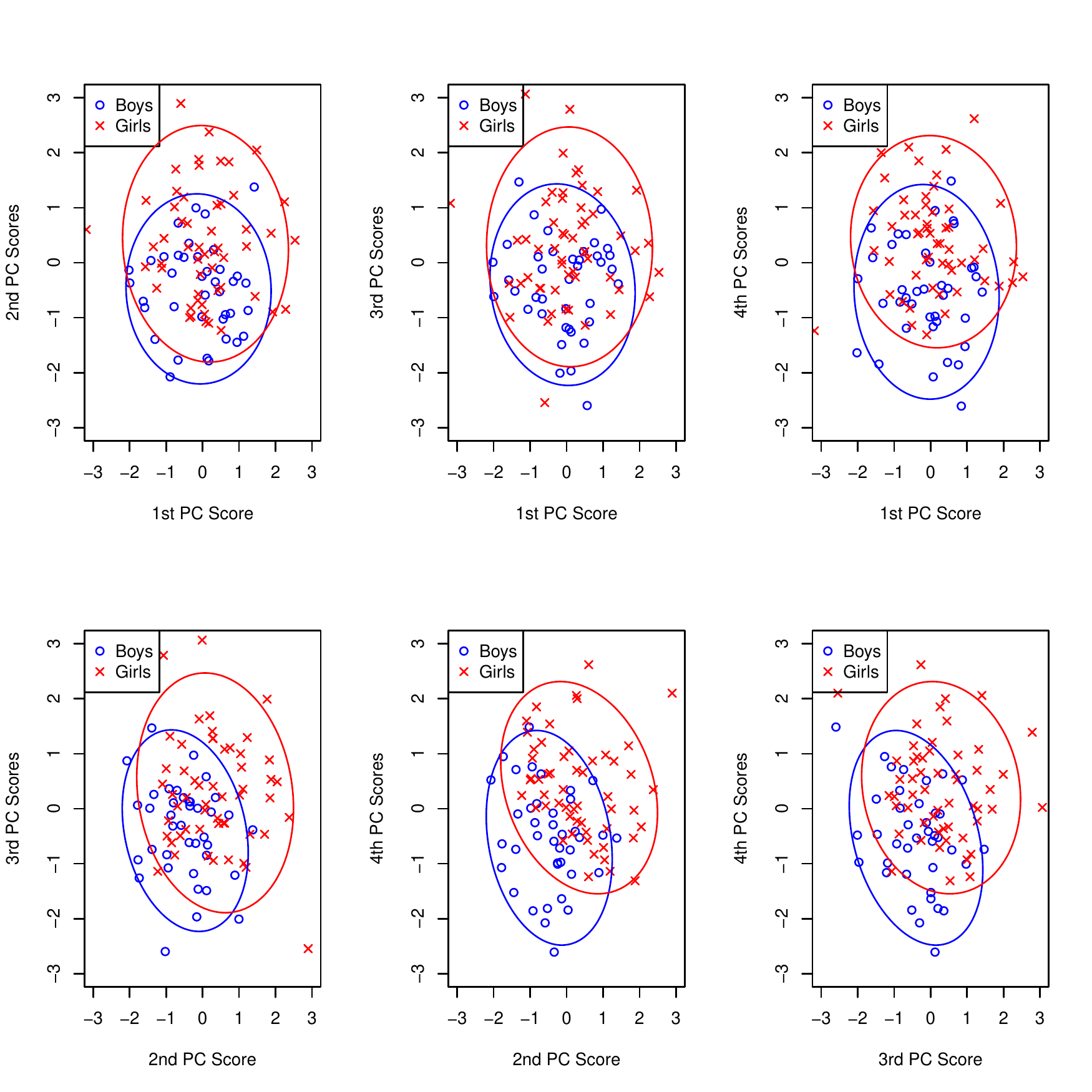}
    \label{fig:PCs1}
  }
  \vfil
  \subfloat[Model 2]{
    \includegraphics[width=0.53\textwidth]{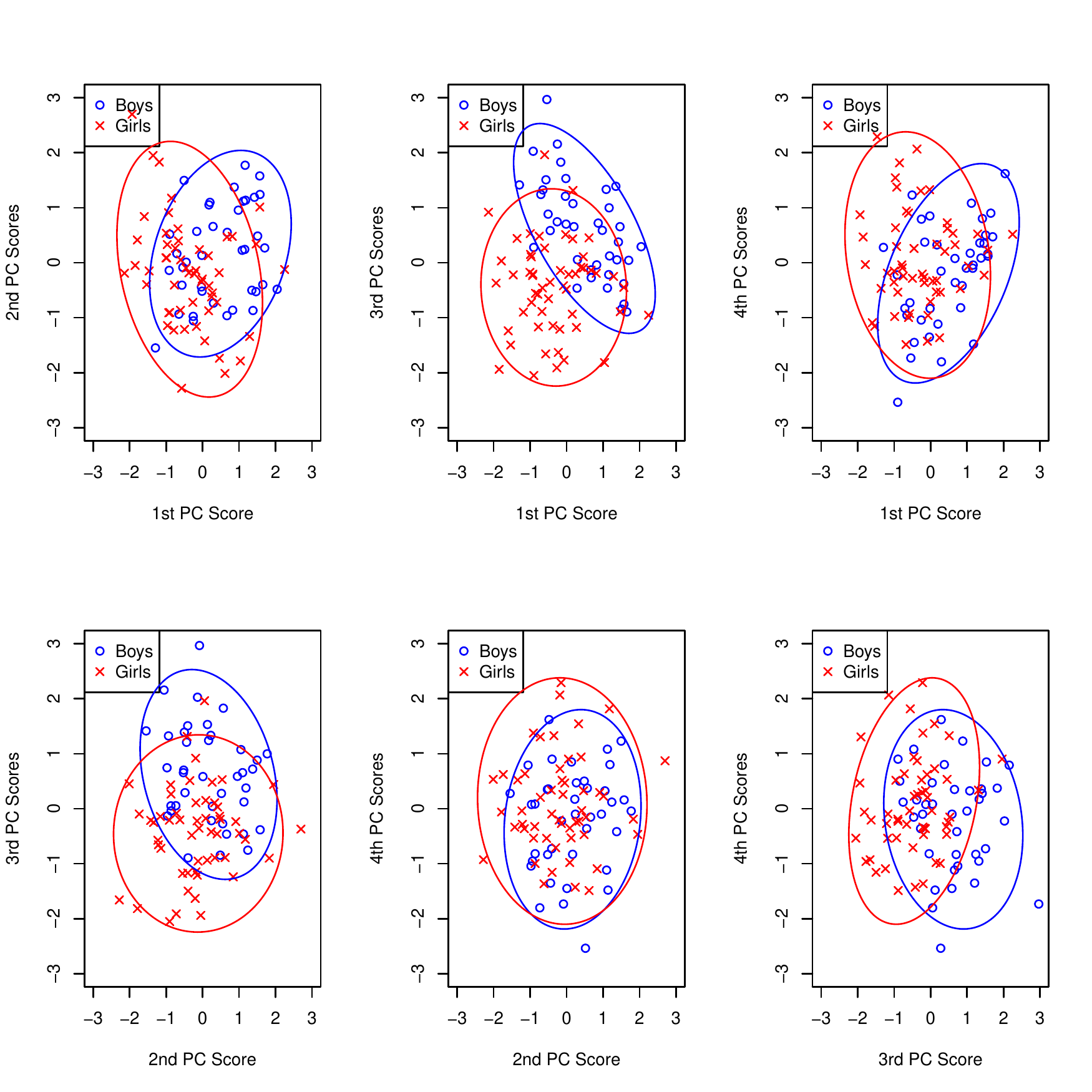}
    \label{fig:PCs2}
  }
  \caption{Scatterplots for the combinations of the standardized PC scores.
    The circles indicate .95th normal-quantile contours transformed by
    the sample mean and variance-covariance matrix.}
  \label{fig:PCs}
\end{figure}

\begin{figure}[h]
  \centering
  \subfloat[Model 1]{
    \includegraphics[width=0.45\textwidth]{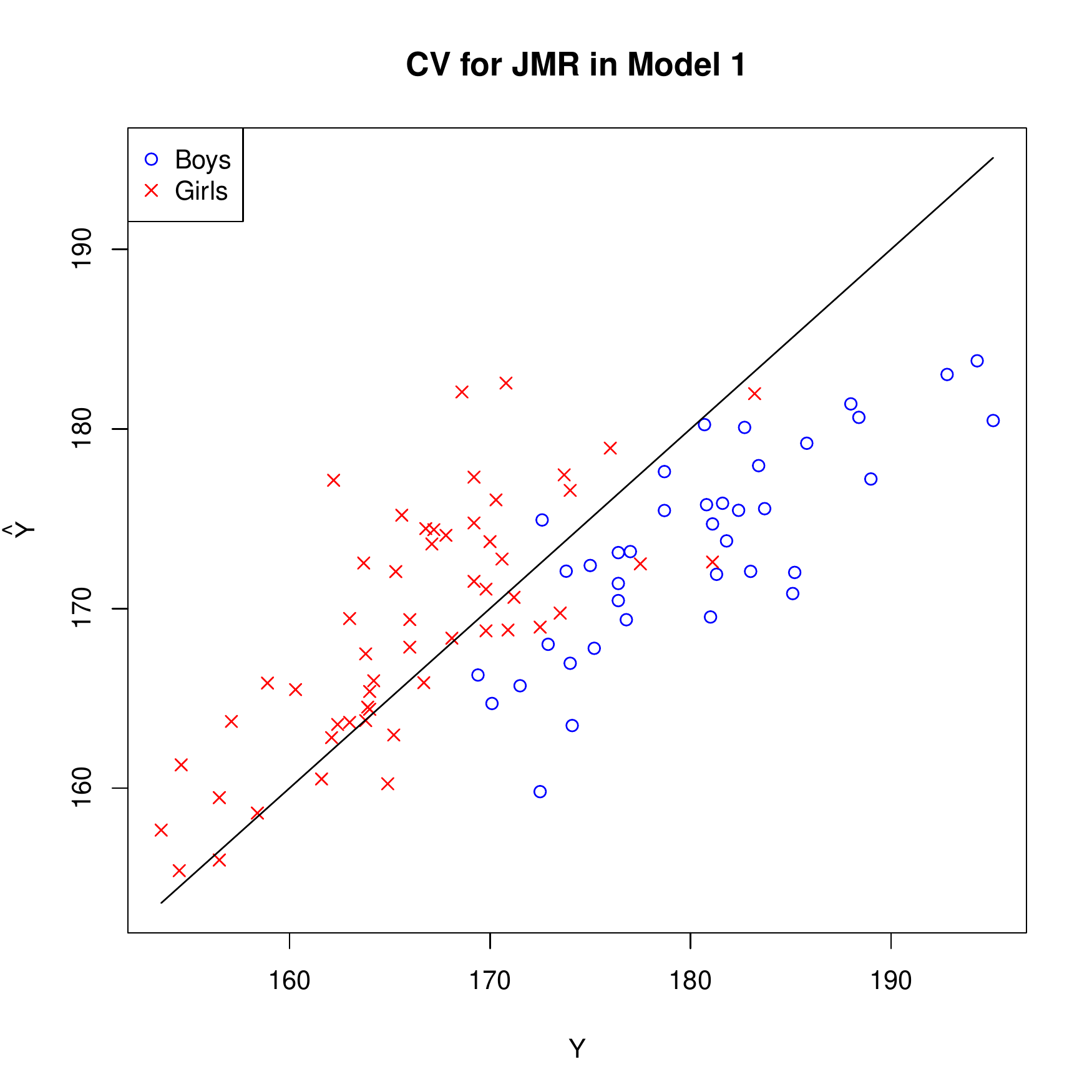}
    \label{fig:CV1}
  }
  \qquad
  \subfloat[Model 2]{
    \includegraphics[width=0.45\textwidth]{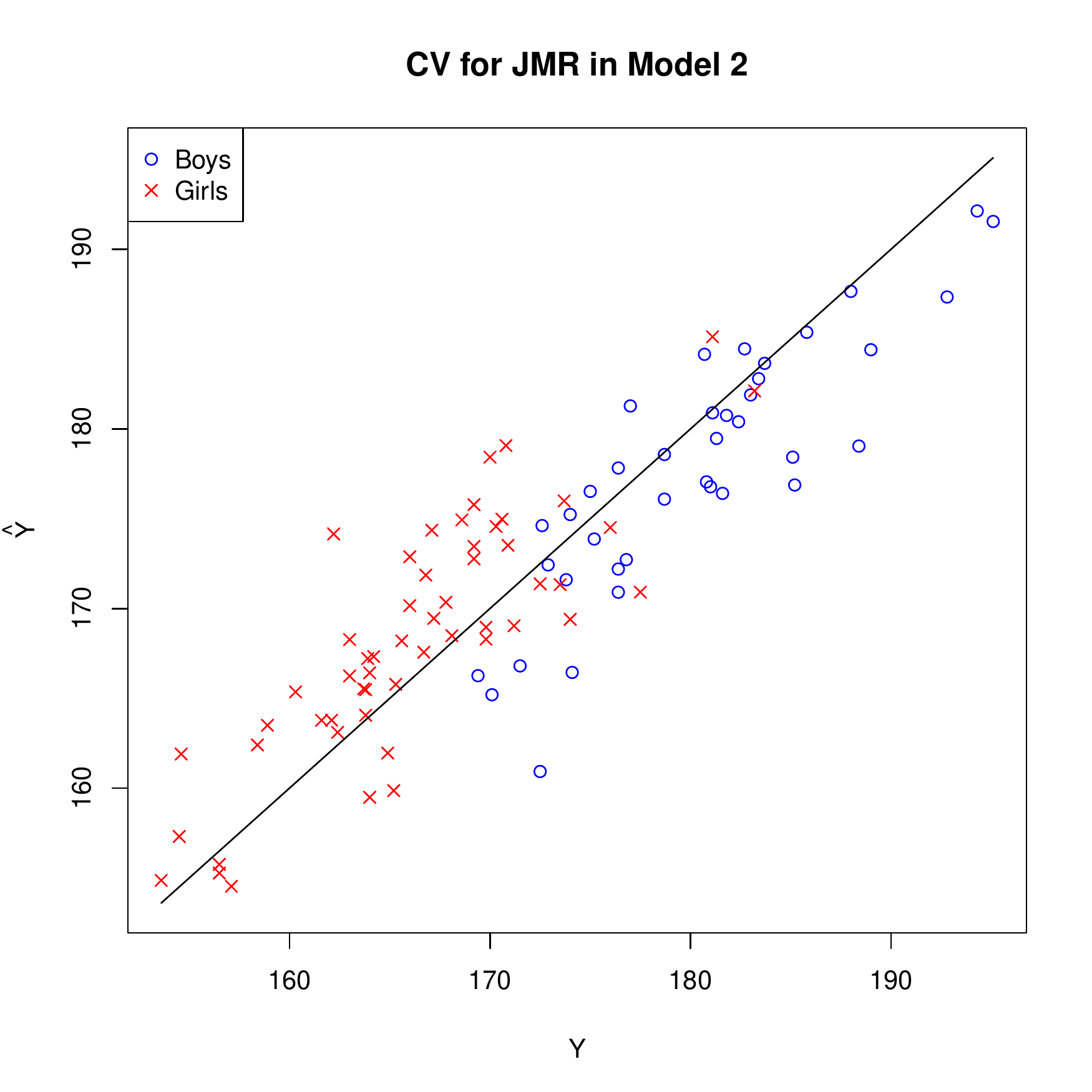}
    \label{fig:CV2}
  }
  \caption{The cross-validated predictors from the one-curve-out samples using joint mixture regression 
    with three leading eigenfunctions.}
  \label{fig:CV}
\end{figure}

We first consider the model where the predictor is a growth curve from 3--12 years old
(see Figure \ref{fig:gcurve}), which is the model that \citet{YFL11} considered (referred as Model 1).
This age period usually contains female pubertal growth peaks near the end of the range;
male pubertal growth peaks usually come several years later.
Given the juvenile growth curve of a new subject, we wish to predict the height at his or her age of 18.
\begin{figure}[h]
  \centering
  \subfloat[Age 12]{
    \label{fig:height12}
    \includegraphics[width=0.45\textwidth]{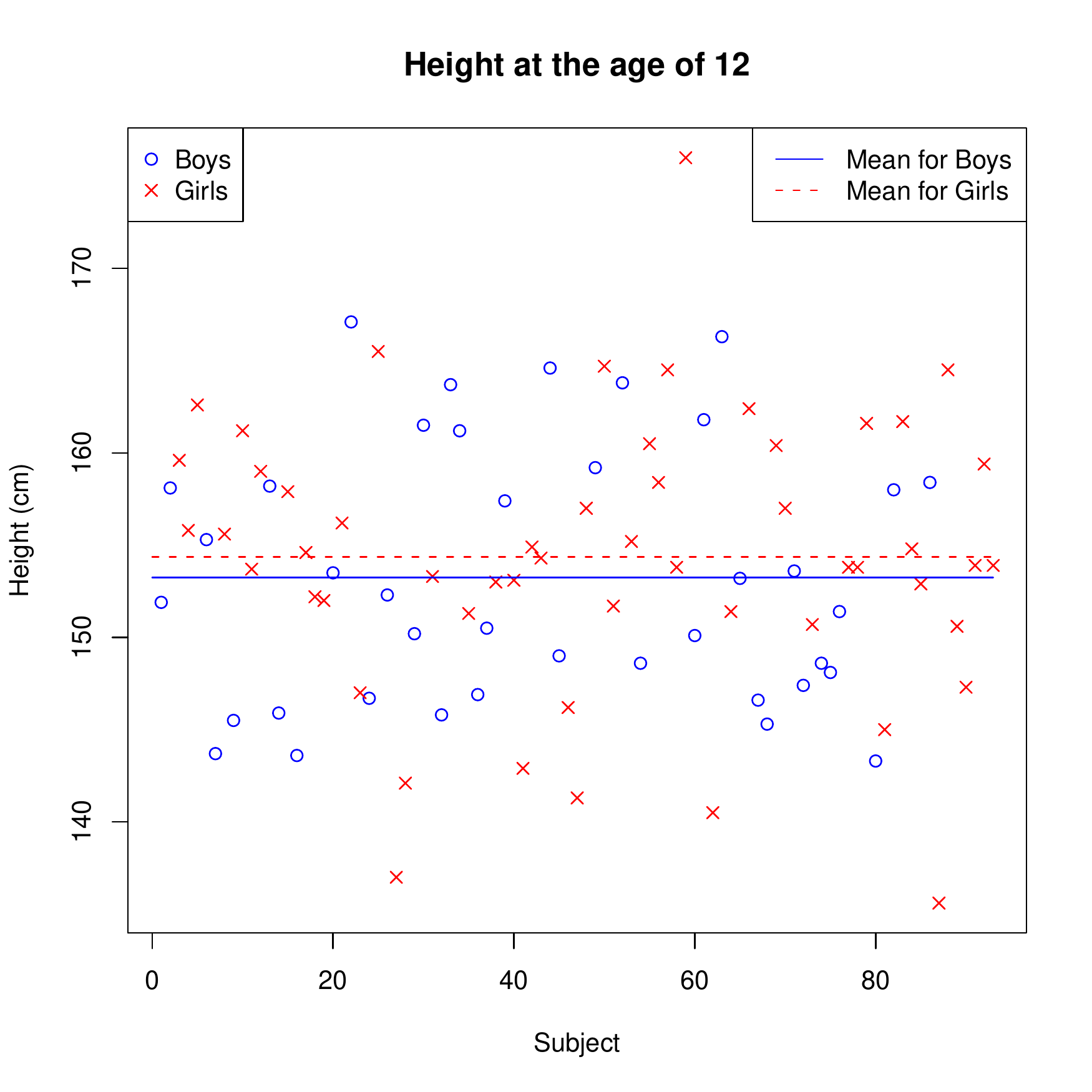}
  }
  \qquad
  \subfloat[Age 18]{
    \label{fig:height18}
    \includegraphics[width=0.45\textwidth]{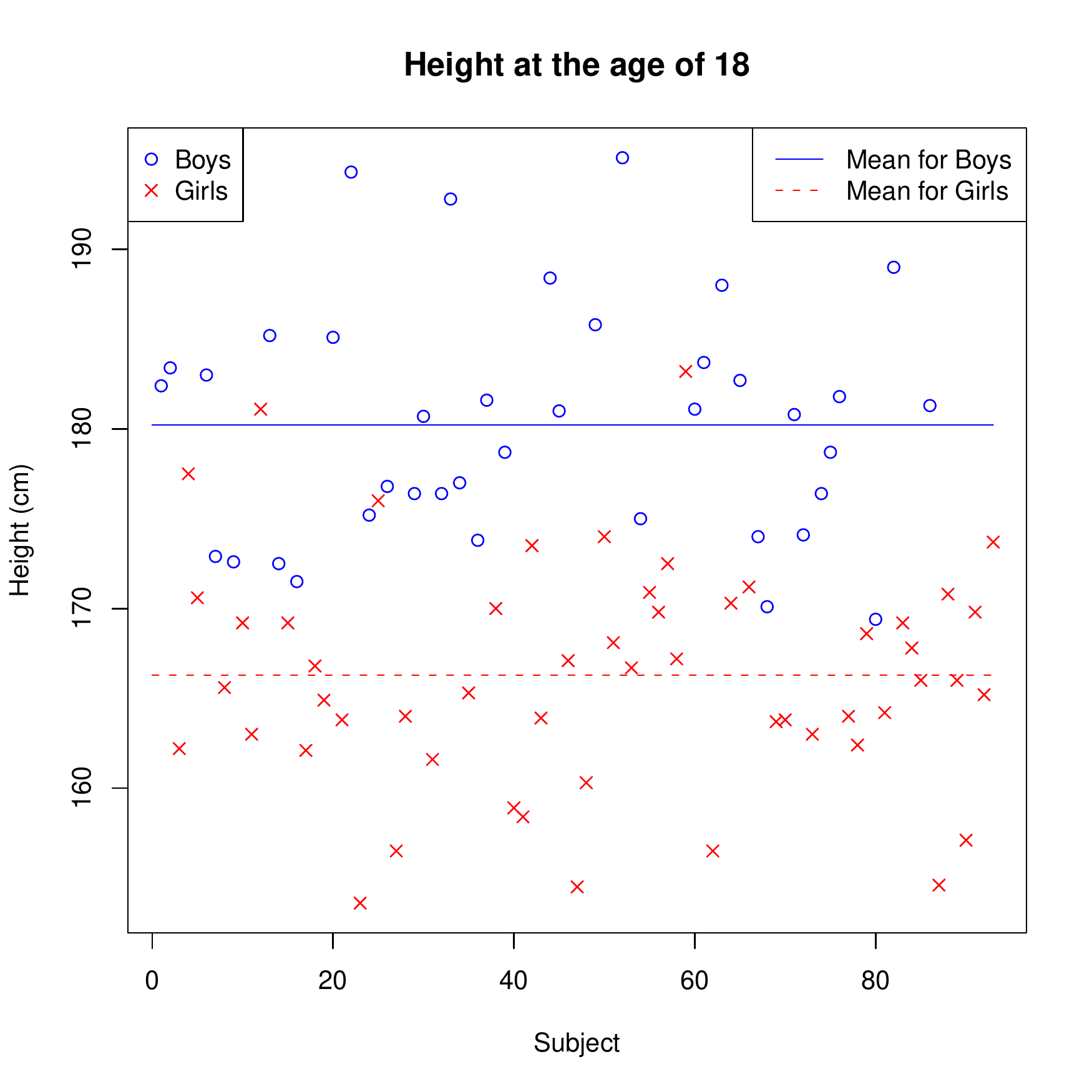}
  }
  \caption{Heights at the age of 12 (Left) and 18 (Right). The order is determined randomly.
  The vertical lines are the sample means for boys and girls.}
  \label{fig:height}
\end{figure}
Figure \ref{fig:height} shows the height at the age of 12 and 18.
It can be seen that predicting the height at the age of 18 from the height at the age of 12 is challenging 
as there is no significant difference in the height distribution at the age of 12 between boys and girls.
Thus to predict the height well it is crucial to differentiate gender from the growth curve; 
recall that we do not assume that gender information is available
(\citet{YFL11} claims that JMR works better than simply fitting a linear regression model, but we suspect that 
they used gender information when predicting the response even though they did not use it when fitting the 
model).
We predict the response by the empirical best predictor (\ref{eq:prediction}).
In addition to OMR and JMR, we also consider functional principal component 
regression (PCR) as an alternative approach for comparison [\citet{CH06}, \citet{HH07}].
PCR estimates the linear model so that it does not cluster a sample.
For these three methods---PCR, OMR, and JMR---we calculated leave-one-curve-out cross validation (CV),
\begin{align*}
  CV = \frac{1}{93}\sum_{i=1}^{93}(Y_{i} - \wh Y_{(-i)})^{2},
\end{align*}
where $\wh Y_{(-i)}$ is calculated by: first estimating the parameters from the entire sample 
except the $i$th subject, and then computing the predictor from $X_{i}$.
\begin{table}[t]
  \centering
  \subfloat[Model 1]{
    \begin{tabular}{lllll}
      \toprule
      \# of eigenfunctions & 2 & 3 & 4 & 5 \\
      \midrule
      PCR & 48.785 & 40.460 & 27.695 & 26.465 \\
      OMR & 53.783 & 47.521 & 27.940 & 28.421 \\
      JMR & 50.412 & 42.369 & 26.618 & 22.901 \\
      CumVar& (0.9857) & (0.9932) & (0.9975) & (0.9993) \\
      \bottomrule
    \end{tabular}
    \label{tab:CVMod1}
  }
  \qquad \vspace{.2in}
  \subfloat[Model 2]{
    \begin{tabular}{lllll}
      \toprule
      \# of eigenfunctions & 2 & 3 & 4 & 5 \\
      \midrule
      PCR & 34.241 & 21.634 & 20.682 & 20.976 \\
      OMR & 36.617 & 21.638 & 22.505 & 23.012 \\
      JMR & 32.888 & 18.134 & 16.929 & 15.293 \\
      CumVar& (0.6584) & (0.7830) & (0.8908) & (0.9882) \\
      \bottomrule
    \end{tabular}
    \label{tab:CVMod2}
  }
  \caption{Cross Validation using the Berkeley Growth Data.
    The last row shows the proportion of the cumulative variance explained by the used eigenfunctions 
    in the total variation (the sum of the eigenvalues).}
  \label{tab:CV}
\end{table}
The results are given in Table \ref{tab:CVMod1};
there are several points that are consistent with what we saw in the simulation study.
Note that JMR displays its advantage over the other methods when using four or more eigenfunctions
while using only two or three eigenfunctions it is not as good as PCR.
Now, looking at Figure \ref{fig:PCs1} where the scatterplots for the estimated standardized principal 
component (PC) scores labeled by gender are shown, it can be seen that the first three PC scores are 
not well separated
by gender while the fourth PC score seems to show some heterogeneity between gender.
Also, looking at Table \ref{tab:MC1}, which shows the number of the misclassification for gender,
it can be seen that JMR clusters the sample by gender very well no matter how many eigenfunctions are used 
while OMR suddenly behaves poorly when using the fourth eigenfunction whose PC score shows the 
differentiability between gender.
These observations support the theory that the prediction performance of JMR depends on the heterogeneity 
of the covariance distribution.
As we saw in the simulation study, OMR performs no better than PCR.
We may wonder if there is a way to improve the model so that JMR performs the best regardless of the number of
the eigenfunctions to be used.
In fact, as seen in Figure \ref{fig:CV1}, which shows the cross-validated predictors from the 
leave-one-curve-out samples using three leading eigenfunctions, JMR suffers from a bias by gender
(most of the male heights locate below the diagonal line while most of the female heights locate above it).
We want JMR to perform in the way that it reduces this group bias by estimating the membership well.
In the second part of this section, we explore an alternative model that uses the tricks we introduced
in Section \ref{sec:incorp}.

\begin{table}[t]
  \centering
  \subfloat[Model 1]{
    \begin{tabular}{lllll}
      \toprule
      \# of eigenfunctions & 2 & 3 & 4 & 5 \\
      \midrule
      OMR & 5 & 6 & 27 & 25 \\
      JMR & 5 & 6 & 6 & 5 \\
      \bottomrule
    \end{tabular}
    \label{tab:MC1}
  }
  \vfil
  \subfloat[Model 2]{
    \begin{tabular}{lllll}
      \toprule
      \# of eigenfunctions & 2 & 3 & 4 & 5 \\
      \midrule
      OMR & 14 & 27 & 31 & 34 \\
      JMR & 8 & 8 & 4 & 6 \\
      \bottomrule
    \end{tabular}
    \label{tab:MC2}
  }
  \caption{The number of misclassifications based on the gender (n=93).}
  \label{tab:MC}
\end{table}

As mentioned in \citet{RS05}, a velocity curve, or an acceleration curve, shows much clearer distinction 
by gender than the original growth curve does.
We incorporate the velocity curve into the model by the way we introduced in Section \ref{sec:incorp}.
In particular, we use the velocity curve from 3--12 years old as the functional covariate and the height at 
the age of 12 as the scalar covariate (referred as Model 2).
We do, however, treat the latter covariate as an invariant covariate since the heights at the age of 12 
for boys and girls are very similar and almost impossible to differentiate (recall Figure \ref{fig:height12}).
Thus it is crucial to estimate gender from the velocity curve to improve the prediction performance.
For PCR and OMR, we simply use these two variables as covariates.
The difference between OMR and JMR under this model is whether we incorporate the distribution of the velocity 
curve into the model.
The results are given in Table \ref{tab:CVMod2}.
First, we notice that the overall prediction performance has dramatically improved from Model 1.
In particular, JMR outperforms the other two approaches regardless of how many eigenfunctions are used.
Looking at Figure \ref{fig:PCs2} where the scatterplots for the estimated standardized PC scores of
the velocity curve are shown, the leading PC scores are much more differentiable by gender than those of
the growth curve (cf. Figure \ref{fig:PCs1}).
Also, Table \ref{tab:MC2} shows that JMR clusters the sample by gender fairly well while OMR no longer do so 
no matter how many eigenfunctions are uses.
OMR again performs no better than PCR.
Note that JMR keeps improving the prediction performance with more eigenfunctions used while PCR and JMR
are stuck at the use of three of four eigenfunctions.
Finally, Figure \ref{fig:CV2} shows that JMR under Model 2 considerably reduces the bias by gender.

Now, we may wonder how large posterior probabilities in the JMR approach actually contribute to improve 
the prediction.
To see this, we calculate CV for the subsamples whose estimated posteriors are larger than 
a certain threshold.
In this analysis, we first estimate the parameters from a leave-one-curve-out sample and compute 
the posteriors (\ref{eq:partpost}) for the subject that is left out.
Then, we compute the mean squared prediction errors by collecting only those subjects whose greater posterior
is larger than a predetermined threshold.
\begin{figure}[h]
  \centering
  \subfloat[Model 1]{
    \label{fig:CV-posterior1}
    \includegraphics[width=0.45\textwidth]{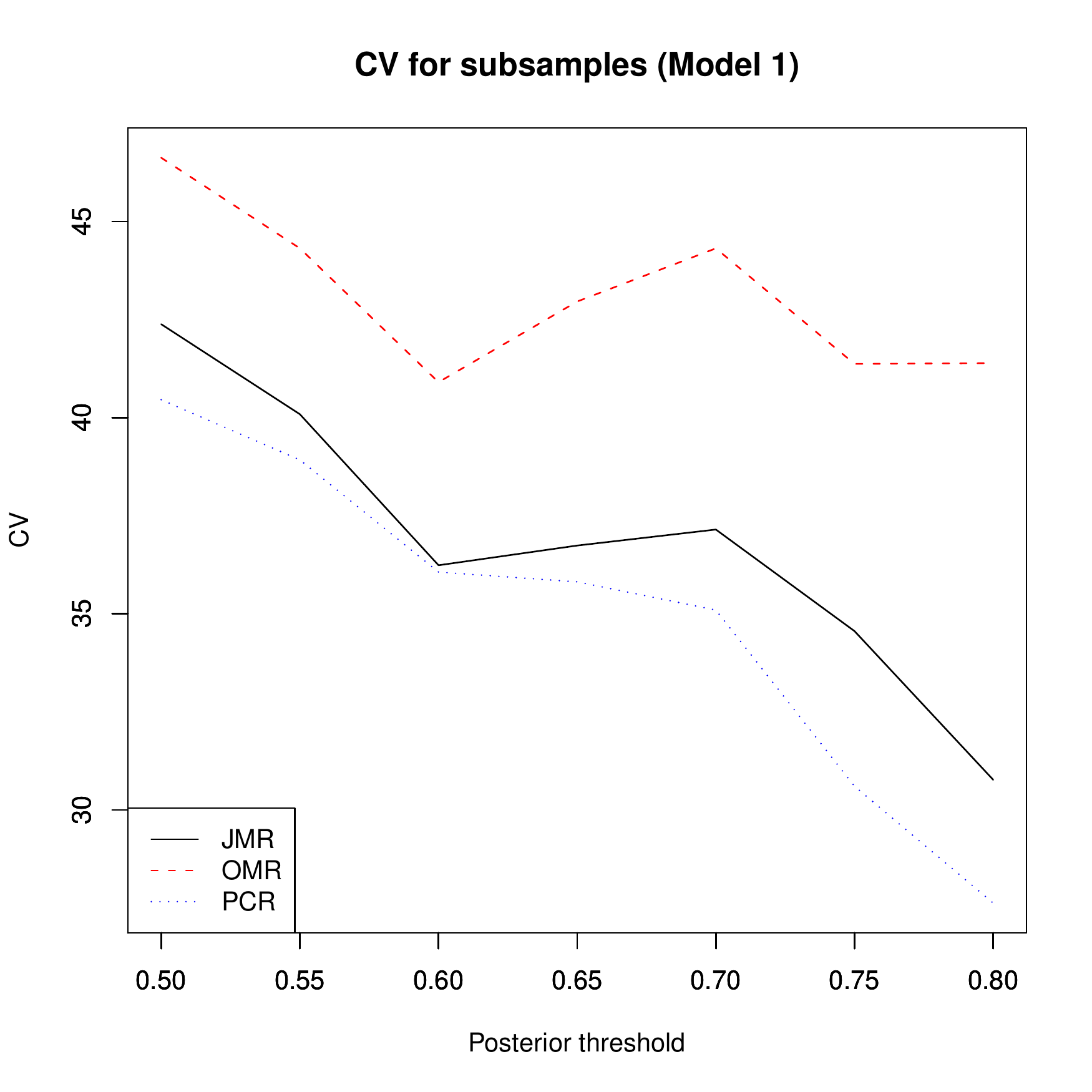}
  }
  \qquad
  \subfloat[Model 2]{
    \label{fig:CV-posterior2}
    \includegraphics[width=0.45\textwidth]{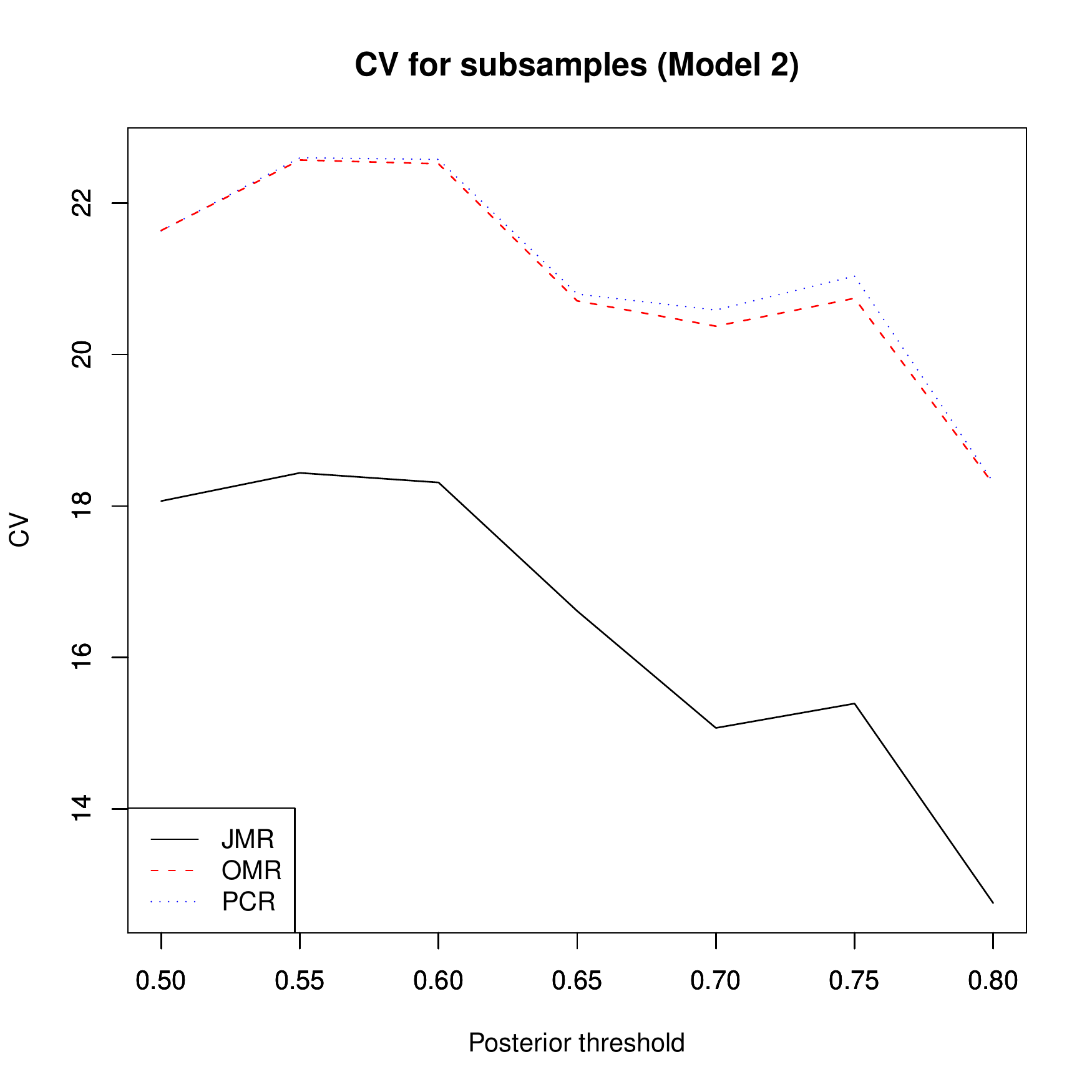}
  }
  \caption{Cross validation comparison for subsamples possessing the estimated posteriors greater than
  or equal to threshold values.}
  \label{fig:CV-posterior}
\end{figure}
Figure \ref{fig:CV-posterior} shows the transition of the CV along different thresholds for the two models
using three leading eigenfunctions.
The thresholds used here are from 0.5 through 0.8 by 0.05 with which the resulting subsample sizes are 
respectively 93, 80, 69, 60, 50, 41, 33 for Model 1 and 93, 86, 84, 79, 73, 69, 59 for Model 2 
(0.5 corresponds to the whole sample).
Overall, Model 2 provides larger posteriors than Model 1 (at each threshold, the subsample size in Model 2
is larger than that in Model 1).
This is consistent with the fact that the three PC scores in Model 2 behaves more differently by gender
than those in Model 1 as seen in Figure \ref{fig:PCs}.
As seen in Figure \ref{fig:CV-posterior2}, in Model 2, JMR improves the prediction performance with a faster
rate than the other two methods as the threshold increases.
In contrast, Figure \ref{fig:CV-posterior1} does not display such behavior;
in fact, PCR performs always better than the other two.
This implies that under Model 2, JMR improves the prediction performance more than the other two 
by estimating the membership from the covariate that is heterogeneous by gender.

\section{Discussion}
\label{sec:discussion}
In this paper, we introduced a mixture regression model where the joint distribution of the response
and the covariate is modeled as a mixture.
We call it joint mixture regression in contrast to the traditional mixture regression,
which we call ordinary mixture regression.
By incorporating the covariate distribution into the model, the heterogeneity of the covariate distribution
across the groups is also taken into account.
From a new observation of the covariate, we can compute the posterior probabilities 
that the subject belongs to each group.
Using these posterior probabilities, the prediction of the response can adaptively use the covariate.
Through the simulation studies and the real-data analysis using the Berkeley growth study data,
we showed that in order to predict the response well, it is crucial that the covariate
behaves differently across the groups.
If the covariate behaves similarly or is deterministic, the mixture regression approach
performs no better than simply fitting a linear regression model.
By including the covariate that behaves differently across the group, we showed that our approach can
significantly improve the prediction performance from the traditional mixture regression approach.

We conclude this paper with two question.
First, as we saw in the simulation study the MLE obtained by fitting the ordinary mixture regression model
may be consistent even under the joint mixture regression model.
We conducted a large number of simulation studies, including the one given in this paper, 
and they all seem to support this conjecture. 
Can we analytically examine the genuineness of this conjecture?
Second, in the functional covariate model we used the eigenfunctions of the observed data as basis functions
onto which the functional covariate is mapped.
However, any basis functions can be used in this procedure.
The best basis functions should be the ones where the projections have the distribution most separable
across the groups so that it becomes easy to estimate the membership from them.
Though using the eigenfunctions of the observed covariate makes an intuitive sense,
analytical justification is lacking.
What basis functions yield the best projection in the joint mixture regression model?
We leave these two questions to be solved in the future.

\newpage
\appendix

\section{Proofs}

\subsection{Consistency of the MLE in the functional covariate model}
\label{sec:proof1}
We need to verify if the likelihood function under the joint mixture regression model behaves appropriately.
Recall that the log-likelihood function is given by
\begin{align*}
  \ell(\Psi; y, \xi) = \log\Big\{\sum_{k=1}^{K}\pi_{k}\varphi(y;a_{k}+b_{k}^{T}\xi,\sigma_{k}^{2})
  \varphi(\xi;\mu_{k},\Sigma_{k})\Big\}.
\end{align*}
The regularity conditions given in \citet{YFL11} are as follows.
For any $\Psi_{1}$ in a pre-fixed compact set defined in the proposition:
\begin{enumerate}
\item [(B1)] There exist some functions $g(y,\xi,\Psi)$ and $c(\Psi)$ such that,
  for all possible values of $y, \xi', \xi''$ and $\Psi\in N_{\Psi_{1}}$,
  where $N_{\Psi_{1}}$ is some neighborhood of $\Psi_{1}$,
  \begin{align*}
    \|\ell(\Psi;y,\xi') - \ell(\Psi;y,\xi'')\| \leq
    g(y,\xi,\Psi)\|\xi'-\xi''\| + c(\Psi)\|\xi'-\xi''\|^{2},
  \end{align*}
  and $g(y,\xi,\Psi)$ and $c(\Psi)$ satisfy
  \begin{align*}
    \sup_{\Psi\in N_{\Psi_{1}}}&\E[g^{2}(y,\xi,\Psi)]< \infty, \\
    \sup_{\Psi\in N_{\Psi_{1}}}&c(\Psi) < \infty,
  \end{align*}
  where the integration is defined by the true parameters.

\item [(B2.1)] $\ell(\Psi;y,\xi)$ is upper semicontinuous in $\Psi\in N_{\Psi_{1}}$ for all $(y,\xi)$.

\item [(B2.2)] There exists a function $D(y,\xi)$ such that $\E D(y,\xi)<\infty$ and
  $\ell(\Psi;y,\xi)\leq D(y,\xi)$ for all $(y,\xi)$ and $\Psi\in N_{\Psi_{1}}$.

\item [(B2.3)] For $\Psi\in N_{\Psi_{1}}$ and sufficiently small $r>0$,
  $\sup_{\Psi':\|\Psi'-\Psi\|<r}q(y,\xi,\Psi')$ is measurable in $(y,\xi)$.
\end{enumerate}
It is easy to see that (B2.1)--(B2.3) are satisfied.
By setting
\begin{align*}
  g(y,\xi,\Psi) &= \sum_{k=1}^{K}\Big[\frac{\|b_{k}\|}{\sigma_{k}^{2}}\{|y-a_{k}|+\|b_{k}\|\|{\xi^{*}}'\|\}
  + \lambda_{\max}(\Sigma_{k})\{\|{\xi^{*}}'\|+\|\mu_{k}\|\}\Big] \\
  c(\Psi) &= \sum_{k=1}^{K}\Big(\frac{\|b_{k}\|^{2}}{\sigma_{k}^{2}}+\lambda_{\max}(\Sigma_{k})\Big),
\end{align*}
where $\lambda_{\max}(\Sigma_{k})$ is the maximum eigenvalue of $\Sigma_{k}$,
(B1) is also satisfied, and all the regularity conditions are satisfied by the likelihood in problem as well.

\subsection{Asymptotic mean squared prediction error} \label{sec:proof2}
We first confirm that $(\ref{eq:mspe2})\geq (\ref{eq:mspe1})$ where the equality holds only when 
$\E(\delta_{k}|X)=\E(\delta_{k})$ for all $k$, or $\la\beta_{1},X\ra = \dots = \la\beta_{K},X\ra$ almost surely.
Denoting $\E(\delta_{k}|X)$ by $p_{k}$ and $\la\beta_{k},X\ra$ by $e_{k}$, the inside of the expectation
operator in $(\ref{eq:mspe2}) - (\ref{eq:mspe1})$ is given by
\begin{align*}
  L := \sum_{k=1}^{K}p_{k}\{\sum_{\ell=1}^{K}\pi_{\ell}(e_{k} - e_{\ell})\}^{2}
  - \sum_{k=1}^{K}p_{k}\{\sum_{\ell=1}^{K}p_{\ell}(e_{k} - e_{\ell})\}^{2}.
\end{align*}
Since $\pi_{K} = 1 - \pi_{1} - \dots - \pi_{K-1}$, for $j=1,\dots, K-1$,
\begin{align*}
  \frac{\partial L}{\partial \pi_{j}} &= 
  2\sum_{k=1}^{K}p_{k}(e_{k} - e_{j})\sum_{\ell=1}^{K}\pi_{\ell}(e_{k} - e_{\ell}) - 
  2\sum_{k=1}^{K}p_{k}(e_{k} - e_{K})\sum_{\ell=1}^{K}\pi_{\ell}(e_{k} - e_{\ell})\\
  &= 2(e_{K} - e_{j})(\sum_{k=1}^{K}p_{k}e_{k} - \sum_{\ell=1}^{K}\pi_{\ell}e_{\ell}),
\end{align*}
which is zero at $\pi_{j}=p_{j}$.
Furthermore,
\begin{align*}
  \frac{\partial^{2}L}{\partial\pi_{j}\partial\pi_{j'}} = 2(e_{K} - e_{j})(e_{K} - e_{j'}),
\end{align*}
thus $[\frac{\partial^{2}L}{\partial\pi_{j}\partial\pi_{j'}}]_{j,j'=1,\dots,K-1}$ is strictly positive definite
unless $e_{1}=\dots=e_{K}$, and the conclusion follow.

We now confirm the other claim.
Note that (\ref{eq:mspe3}) can be rewritten to
\begin{align*}
  \text{I} &:= \sum_{k=1}^{K}\pi_{k}\E[\la\sum_{\ell=1}^{K}\pi_{\ell}^{*}(\beta_{k}-\beta_{\ell})+{\cal B},
  X_{k}\ra^{2}],
\end{align*}
where ${\cal B}:= \sum_{\ell=1}^{K}\pi_{\ell}^{*}(\beta_{\ell}-\beta_{\ell}^{*})$,
while (\ref{eq:mspe2}) can be rewritten to
\begin{align*}
  \text{II} := \sum_{k=1}^{K}\pi_{k}\E\Big[\Big\la\sum_{\ell=1}^{K}\pi_{\ell}(\beta_{k}-\beta_{\ell}),
  X_{k}\Big\ra^{2}\Big]
\end{align*}
We will prove $\text{I} - \text{II} \geq 0$ under the assumption 
$\Gamma=\E(X_{1}X_{1}^{T})=\dots=\E(X_{K}X_{K}^{T})$.
Observe
\begin{align*}
  \text{I} &= \sum_{k=1}^{K}\pi_{k}\E[\la\sum_{\ell=1}^{K}\pi_{\ell}^{*}(\beta_{k} - \beta_{\ell})+{\cal B},
  X_{k}\ra^{2}]\\
  &= \sum_{k=1}^{K}\pi_{k}\E[\la(\beta_{k} - \beta_{K}) +
  \sum_{\ell=1}^{K-1}\pi_{\ell}^{*}(\beta_{K} - \beta_{\ell}) + {\cal B}, X_{k}\ra^{2}] \\
  &= \sum_{k=1}^{K}\pi_{k}\{(\beta_{k} - \beta_{K}) +
  \sum_{\ell=1}^{K-1}\pi_{\ell}^{*}(\beta_{K} - \beta_{\ell})\}^{T}
  \Gamma \{(\beta_{k} - \beta_{K}) + \sum_{\ell=1}^{K-1}\pi_{\ell}^{*}(\beta_{K} - \beta_{\ell})\} \\
  &\qquad + 2{\cal B}^{T}\Gamma\sum_{k=1}^{K}\pi_{k}\{(\beta_{k} - \beta_{K}) +
  \sum_{\ell=1}^{K-1}\pi_{\ell}^{*}(\beta_{K} - \beta_{\ell})\} + {\cal B}^{T}\Gamma{\cal B}.
\end{align*}
Similarly,
\begin{align*}
  \text{II} &= \sum_{k=1}^{K}\pi_{k}\E[\la\sum_{\ell=1}^{K}\pi_{\ell}(\beta_{k} - \beta_{\ell}),X_{k}\ra^{2}] \\
  &= \sum_{k=1}^{K}\pi_{k}\{(\beta_{k} - \beta_{K}) + \sum_{\ell=1}^{K-1}\pi_{\ell}(\beta_{K} - \beta_{\ell})\}^{T}
  \Gamma \{(\beta_{k} - \beta_{K}) + \sum_{\ell=1}^{K-1}\pi_{\ell}(\beta_{K} - \beta_{\ell})\}
\end{align*}
Observe that the first term of I minus II is given by
\begin{align*}
  &-2\{\sum_{\ell=1}^{K-1}\pi_{\ell}^{*}(\beta_{K}-\beta_{\ell})\}^{T}\Gamma
  \{\sum_{\ell=1}^{K-1}\pi_{\ell}(\beta_{K}-\beta_{\ell})\}^{T} + 
  \{\sum_{\ell=1}^{K-1}\pi_{\ell}^{*}(\beta_{K}-\beta_{\ell})\}^{T}\Gamma
  \{\sum_{\ell=1}^{K-1}\pi_{\ell}^{*}(\beta_{K}-\beta_{\ell})\}^{T} \\
  &+2\{\sum_{\ell=1}^{K-1}\pi_{\ell}(\beta_{K}-\beta_{\ell})\}^{T}\Gamma
  \{\sum_{\ell=1}^{K-1}\pi_{\ell}(\beta_{K}-\beta_{\ell})\}^{T} -
  \{\sum_{\ell=1}^{K-1}\pi_{\ell}(\beta_{K}-\beta_{\ell})\}^{T}\Gamma
  \{\sum_{\ell=1}^{K-1}\pi_{\ell}(\beta_{K}-\beta_{\ell})\}^{T} \\
  &\qquad = \{\sum_{\ell=1}^{K-1}(\pi_{\ell}^{*}-\pi_{\ell})(\beta_{K}-\beta_{\ell})\}^{T}\Gamma
  \{\sum_{\ell=1}^{K-1}(\pi_{\ell}^{*}-\pi_{\ell})(\beta_{K}-\beta_{\ell})\}.
\end{align*}
The second term of I can be rewritten as
\begin{align*}
  2{\cal B}^{T}\Gamma\sum_{\ell=1}^{K-1}(\pi_{\ell}^{*}-\pi_{\ell})(\beta_{K}-\beta_{\ell}).
\end{align*}
Thus we have
\begin{align*}
  \text{I} - \text{II} 
  = \{\sum_{\ell=1}^{K-1}(\pi_{\ell}^{*}-\pi_{\ell})(\beta_{K}-\beta_{\ell}) + {\cal B}\}^{T}\Gamma
  \{\sum_{\ell=1}^{K-1}(\pi_{\ell}^{*}-\pi_{\ell})(\beta_{K}-\beta_{\ell}) + {\cal B}\} \geq 0.
\end{align*}


\section{The EM algorithm for joint mixture regression}\label{sec:em}
Denote the data matrix by $X = [x_{1},\dots, x_{n}]^{T}$, and let $\ol X = [\ol x_{1},\dots,\ol x_{n}]^{T}$ 
where $\ol x_{i} = [1,x_{i}^{T}]^{T}$, so that $\ol X$ is a $n\times (1+p)$ matrix.
Also, let $\ol\beta_{k} = [\alpha_{k},\beta_{k}^{T}]^{T}$.
Once the M-step is done, the next E-step is given by
\begin{align*}
  \tau_{ik}^{(+)} = \frac{\wh\pi_{k}\varphi(y_{i}; \la\wh{\ol\beta}_{k}, \ol x_{i}\ra, \wh\sigma_{k}^{2})
    \varphi(\ol x_{i}; \wh\mu_{k},\wh\Sigma_{k})}{\sum_{k=1}^{K}\wh\pi_{k}
    \varphi(y_{i}; \la\wh{\ol\beta}_{k},\ol x_{i}\ra, \wh\sigma_{k}^{2})
    \varphi(\ol x_{i};\wh\mu_{k},\wh\Sigma_{k})},
\end{align*}
where the inner product is the usual inner product in $\mathbb R^{p+1}$ and 
the hat denotes the estimate obtained in the last M-step.

The M-step is conducted as follows. Define for $k=1,\dots,K$,
\begin{align*}
  \wh W_{k} &= \diag\{\wh\tau_{1k},\dots, \wh\tau_{nk}\}, \quad \wt X_{k} = \wh W_{k}^{1/2}X,
  \quad \wt{\ol X_{k}} = \wh W_{k}^{1/2}\ol X, \\
  \mathbb 1 &= [1,\dots, 1]^{T} \in \mathbb R^{n}, \quad \wt{\mathbb 1} = \wh W_{k}^{1/2}\mathbb 1, \\
  \wt y_{k} &= \wh W_{k}^{1/2}y, \quad y = [y_{1}, \dots, y_{n}]^{T} \\
  H(\wt{\mathbb 1}) &= \wt{\mathbb 1}(\wt{\mathbb 1}^{T}\wt{\mathbb 1})^{-1}\wt{\mathbb 1}^{T}, \quad
  H(\wt{\ol X_{k}}) = \wt{\ol X_{k}}(\wt{\ol X_{k}}^{T}\wt{\ol X_{k}})^{-1}\wt{\ol X_{k}}^{T},
\end{align*}
where $\wh\tau_{ik}$ are the estimates obtained in the last E-step.
Then, the new M-step is given by
\begin{align*}
  \mu_{k}^{(+)} &= \{(\wt{\mathbb 1}^{T}\wt{\mathbb 1})^{-1}\wt{\mathbb 1}^{T}\wt{X_{k}}\}^{T}, \quad
  \Sigma_{k}^{(+)} = (\wt{\mathbb 1}^{T}\wt{\mathbb 1})^{-1}\wt{X_{k}}^{T}\{I-H(\wt{\mathbb 1})\}\wt{X_{k}}, \\
  \pi_{k}^{(+)} &= \frac{1}{n}\sum_{i=1}^{n}\wh\tau_{ik}, \quad
  \ol\beta_{k}^{(+)} = (\wt{\ol X_{k}}^{T}\wt{\ol X_{k}})^{-1}\wt{\ol X_{k}}^{T}\wt y_{k}, \quad
  \wh \sigma_{k}^{2(+)} = (\wt{\mathbb 1}^{T}\wt{\mathbb 1})^{-1}\wt y_{k}^{T}\{I-H(\wt{\ol X_{k}})\}\wt y_{k},
\end{align*}
for $k=1,\dots,K$.

\newpage
\bibliographystyle{abbrvnat}
\bibliography{mixture,fda}

\end{document}